\documentclass[iop,apj]{emulateapj}

\usepackage{graphicx}
\usepackage{epsf}
\usepackage{natbib}
\usepackage{amsmath}
\usepackage{apjfonts}
\usepackage{color}
\usepackage{times}
\usepackage{microtype}
\usepackage[T1]{fontenc}
\usepackage{pslatex}
\newcommand \lta {\mathrel{\vcenter
     {\hbox{$<$}\nointerlineskip\hbox{$\sim$}}}}
\newcommand \gta {\mathrel{\vcenter
     {\hbox{$>$}\nointerlineskip\hbox{$\sim$}}}}

\newcommand \rr {r_\odot}

\newcommand\kms{km~s$^{-1}$}
\newcommand\gcm{g~cm$^{-3}$}

\newcommand\degrees{^{\rm o}}
\newcommand\waphi{\omega_{\rm {A}\phi}}
\def\gtsim{\lower.5ex\hbox{$\buildrel > \over\sim$}}
\def\ltsim{\lower.5ex\hbox{$\buildrel < \over\sim$}}
\newcommand\kdotva{\mathbf{k\cdot v_{\rm A}}}
\begin{document}

\title{The Role of the Magnetorotational Instability in the Sun}

\author{Daniel Kagan, J. Craig Wheeler}
\affil{Department of Astronomy, University of Texas, Austin, TX 78712}
\email{kagan@astro.as.utexas.edu, wheel@astro.as.utexas.edu}

\begin{abstract}
We calculate growth rates for nonaxisymmetric instabilities including the magnetorotational instability (MRI) throughout the Sun. We first derive a dispersion relation for nonaxisymmetric instability including the effects of shear, convective buoyancy, and three diffusivities (thermal conductivity, resistivity, and viscosity). We then use a solar model evolved with the stellar evolution 
code MESA and angular velocity profiles determined by Global Oscillations Network Group (GONG) helioseismology to determine the unstable modes present at each location in the Sun and the associated growth rates. The overall instability has unstable modes throughout the convection zone and also slightly below it at middle and high latitudes. It contains three classes of modes: large--scale hydrodynamic convective modes, large--scale hydrodynamic shear modes, and small--scale magnetohydrodynamic (MHD) shear modes, which may be properly called MRI modes.  While large--scale convective modes are the most rapidly growing modes in most of the convective zone, MRI modes are important in both stably stratified and convectively unstable locations near the tachocline at colatitudes $\theta<53 \degrees$.  Nonaxisymmetric MRI modes grow faster than the corresponding axisymmetric modes; for some poloidal magnetic fields, the nonaxisymmetric MRI growth rates are similar to the angular rotation frequency $\Omega$, while axisymmetric modes are stabilized. We briefly discuss the saturation of the field produced by MRI modes, finding that the implied field at the base of the convective zone in the Sun is comparable to that derived based on dynamos active in the tachocline and that the saturation of field resulting from the MRI may be of importance even in the upper convection zone.  

\end{abstract}

\section{Introduction}
\label{sec:intro}
One of the major unsolved problems of stellar evolution
is the effect of rotation, especially differential rotation, on the magnetic field structure of stars, and
the feedback of that magnetic field on the stellar structure
and evolution \citep[and references therein]{M09}. While the detailed interior magnetic field structure is not known for any star, Global Oscillations Network Group  \citep[GONG, ][]{Ho05} and  Michelson Doppler Imager  \citep[MDI, ][]{CD96} helioseismology  has allowed the detailed calculation of the rotation profile of the Sun. In the radiative zone at $r<0.69~\rr$, the Sun has approximately solid--body rotation. In the convection zone at $r>0.72~\rr$ the rotation rate can be primarily described as a slowly increasing function of spherical $\theta$, except near the solar surface at $r>0.93~\rr$ where a strong radial shear layer is present with $\partial \ln \Omega /\partial \ln r <0$. The radius of the transition from radiative to convective transport is $r\approx0.713 ~\rr$, and the transition is associated with a very strong radial shear layer with width $\sim0.02~\rr$ known as the tachocline, which has $\partial \ln \Omega /\partial \ln r <0$ close to the poles and $\partial \ln \Omega \partial \ln r >0$ close to the equator. The tachocline's central radius is $r\sim0.70~\rr$, and varies slightly with latitude; it is likely located entirely in the radiative region near the equator, but it may reach significantly into the convective region closer to the poles \citep{BA01}.   Combining this observed rotation profile with observed solar surface magnetic fields promises significant progress in understanding the origin of interior solar magnetic fields. 

Observations of surface solar magnetic fields indicate the presence of fields on large spatial scales that vary with the solar cycle and are associated with active regions and eruptive events. The most popular approach to understanding the origin of these large--scale fields involves $\alpha$--$\Omega$ dynamo models associated with the tachocline \citep[e.g.,][and references therein]{OD03}.  In these models, nonaxisymmetric instability or turbulence produces relatively small--scale, nonaxisymmetric poloidal fields from toroidal fields in the lower convection zone and convective overshoot region of the upper tachocline (the $\alpha$ effect).  Field line wrapping by strong differential rotation (the $\Omega$ effect) in the tachocline then stretches poloidal fields into large--scale toroidal fields and completes the dynamo loop. The process behind the $\Omega$ effect which produces toroidal fields by stretching poloidal fields is well understood, but there are many candidate mechanisms for the $\alpha$ effect that produces poloidal fields from toroidal fields. One possible way of producing an $\alpha$ effect is a kinematic mean--field dynamo based on the \citet{p55} mechanism, in which small--scale, convective, nonaxisymmetric turbulence causes toroidal fields to be passively advected into poloidal fields. Passive advection is also used to produce poloidal fields in Babcock--Leighton models, in which the source of the $\alpha$ effect is the twisting of large--scale toroidal field structures under the influence of the coriolis force as they rise through the convection zone.   An alternative possibility for the $\alpha$ effect is that a local or global hydrodynamic or magnetohydrodynamic (MHD) instability in the tachocline region is directly responsible for the production of poloidal field. A global hydrodynamic instability may be present in the tachocline as a result of the latitudinal differential rotation \citep{dikpati_analysis_2001}. Local MHD instabilities that can be driven by strong radial shear in the tachocline include the Tayler--Spruit instability \citep{pt85,spruit02} and the magnetorotational instability (MRI), which is the subject of this paper.  

In addition to the large--scale fields associated with solar activity, small--scale fields have been observed on the quiet sun that do not vary with the solar cycle. These fields are typically thought to result from small--scale kinetic dynamo action due to convective turbulence \cite[e.g., ][and references therein]{pillet_solar_2013}, probably {\sl in situ} in the upper convection zone \citep{buehler_quiet_2013}.  The presence of a strongly radial shear near the surface of the Sun indicates that MHD instabilities like the MRI and the Tayler--Spruit instability may also grow in the convective zone and play a role in the origin of these magnetic fields.

\subsection{MRI}

As a star, the Sun is subject to a wide variety of thermal, dynamic, and magnetohydrodynamic instabilities, including convective thermal instabilities, baroclinic wind shear instabilities, Kelvin--Helmholtz shear instabilities, and magnetic buoyancy instabilities. These processes, and others, work in concert to establish the physical state of the Sun. In practice, it is often necessary to address various physical effects in relative isolation to elucidate their significance. In this work, we concentrate on the magnetorotational instability \citep[MRI;][]{Vel59, chandra60, 
acheson78, BH91, BH98}. The MRI has been thoroughly explored in the context of accretion disks,but
it also applies to quasi--spherical objects, 
e.g. stars \citep{BH94}. A general dispersion relation and associated instability criteria encompassing the MRI and other instabilities for the 
nonaxisymmetric, diffusive case with finite resistivity, viscosity, 
and conductivity, is presented in \citet{acheson78}. That is the 
relation that should be applied in stars, but it remains cumbersome 
to employ and relatively unexplored. Therefore, other dispersion relations have been derived that include small--scale magnetohydrodynamic (MHD) shear modes that have similar properties to the MRI modes in accretion disks, as well as other large--scale hydrodynamic modes associated with convection and shear. These dispersion relations typically neglect the effects of field gradients and hence of magnetic buoyancy, but include the axisymmetric Kelvin--Helmholtz instability in a context without sharp boundary layers. In Section \ref{sec:otherinstabilities}, we discuss situations in the Sun where the Kelvin--Helmholtz instability may be active \citet{BH94} \citep[see also][]{Ch79} derived a dispersion relation including the MRI for axisymmetric, non--diffusive conditions with rotation restricted to cylinders.  \citet{bal95} generalized the dispersion relation to a general distribution of angular velocity, $\Omega(\varpi, Z)$, where $\varpi$ is the cylindrical radius and $Z$ is the vertical coordinate, while \citet{MBS04} derived a dispersion relation including the effects of viscous, thermal, and magnetic diffusivities.    \citet{KO00} and \citet{MST06} considered the nonaxisymmetric, non--diffusive modes including the MRI in the context of shearing winds and proto--neutron stars, respectively; the latter also included possible effects of magnetic buoyancy modes. \citet{MSS07} explored a nonaxisymmetric dispersion relation containing the MRI including the effects of the three classical diffusivities and neutrino diffusion in the context of proto--neutron stars with spherically symmetric rotation profiles.

\subsection{MRI in the Sun}

To zeroth order, the MRI is driven by a negative gradient of angular velocity, $\Omega$, in a gravitating object. The remarkable feature of this instability is that it depends only on the shear and not on the amplitude of the magnetic field. Other magnetic instabilities, such as the buoyancy instability (Mizerski et al 2013) or the Tayler--Spruit instability (Spruit 2002), manifestly do depend on the strength of the magnetic field. In practice, these magnetic instabilities are likely to act together in the Sun. The MRI acts as a dynamo and can grow small fields to finite strength at saturation, at which point they may be subject to, e.g., buoyancy instabilities driven by gradients in the field strength. In this paper we ignore initial gradients in the magnetic field as we estimate instability to the MRI. Because of the strong shear at the tachocline (Spiegel \& Zahn 1992), the MRI alone could generate fields that are carried by buoyancy to the surface with observed field strength. We return to this point in the conclusions. A principal goal of the current work is to reemphasize that the MRI represents important physics that should be considered in complete models of the Sun that could supplement or complement $\alpha$--$\Omega$ dynamos, interface dynamos, and variations on that theme. Numerical work has elucidated the role of global magnetic instabilities in the Sun and stars (Gilman \& Fox,1997; Brown et al. 2011; Dikpati 2012), but such global models are unable to resolve the wavelengths of the most rapidly growing MRI models and so omit the physics we address here. A more complete solution would incorporate the local instability to the MRI in a global context, a task still challenging for current computational resources and beyond the scope of this paper. Ideally, one would begin this problem with a self--gravitating cloud of proto--stellar gas threaded with a small ambient magnetic field. Given the impracticality of that task, we are forced to make some assumptions and to adopt certain constraints from observations of the current Sun. The current rotational structure of the Sun is the solution of a complex gravo--magnetic interaction. In this work, we adopt the rotational profile of the Sun as determined by studies of solar oscillations. We assume a low fiducial initial magnetic field strength to emphasize that the instability does not depend on the field strength. We have not attempted to address the current internal field structure of the Sun, since that is not well known, but some of the initial field parameters we have chosen correspond to fields comparable to those deduced for the current Sun (Figure 10 corresponds to poloidal fields of $\lta 10$ G and toroidal fields of $\lta 5000$ G).

   The most rapidly growing modes of the MRI typically have $k v_{\rm A}\sim \Omega$, where $v_{\rm A}\equiv B/\sqrt{4 \pi \rho}$ is the Alfv\'en speed,  $k$ is the poloidal wavenumber, $B$ is the initial magnetic field and $\rho$ is the density. The length scale of these modes is therefore $\lambda\sim  2 \pi v_{A}/{\Omega}$. For a seed field $B \sim 1~ G $ in the tachocline, where roughly $\rho\sim0.1 ~{\rm g \ {cm}^{-3}}$  and $\Omega\sim10^{-6}$~rad s$^{-1}$, we obtain $\lambda\sim 10^5 ~{\rm cm} \sim 10^{-5} ~\rr$. Because the most rapidly growing modes of the MRI are so small relative to the solar radius, it is currently impractical to resolve the MRI in global simulations of the Sun. As a result, most analysis of the MRI in stars has been confined to application of local MRI dispersion relations. The action in the Sun of the triply-diffusive axisymmetric instability derived by \citet{MBS04} has been investigated by \citet{MBS04} in the radiative region, by \citet{PM07} in the solar tachocline, and by \citet{M11} in the tachocline and convective zone in isolation from the effects of convection. \citet{MBS04} note that in their extensive stability analysis in 
both the inviscid and perfect--conductor double--diffusive limits
\textit{any} level of negative differential rotation is destabilized by a 
combination of diffusion--free (along spherical shells) and 
double--diffusive (across spherical shells) modes.  They found, however, that even a relatively small viscosity
could add some stability in the triply--diffusive case for differential
rotation between shells, $d \ln \Omega/d \ln \varpi < 0$, so that this case, which is the most 
important one in the Sun, must be considered quantitatively.
 
\citet{PM07} investigated the growth rate of modes in the stably stratified tachocline, where only MRI modes of the dispersion relation derived by \citet{MBS04} exist. Using a semianalytical prescription 
for the differential rotation along the tachocline, they found that the 
regions of $\theta \lta 60 \degrees$ are formally unstable to the MRI, but significant growth of instability occurs only for $\theta < 53 \degrees$. Parfrey \& Menou
concluded that the turbulence associated with the MRI at high latitudes 
disrupts the formation of large scale magnetic fields. They argued that 
such large scale magnetic fields can only form at lower latitudes by more
traditional solar dynamos operating in the tachocline. 

\citet{M11} employed the triply--diffusive dispersion relation 
for the axisymmetric instability of \citet{MBS04} combined with rotation profiles 
determined from helioseismology and a standard model of the Sun to 
calculate the growth rate of the MRI throughout the Sun. He neglected the 
destabilizing effects of thermal buoyancy, but included the effects of 
stable stratification, excluding all non--MRI modes from the analysis. He found that unstable modes existed in the tachocline at the same latitudes found by \citet{PM07} as well as near the solar surface at low latitudes $\theta> 45 \degrees$. He argued that the calculated growth of instability at large radii near the solar surface was unrealistic, because the strong convection present there would disrupt this growth. 

  In this paper, we consider the full triply--diffusive, nonaxisymmetric
magneto--convective instability in a model of the current Sun; we thus assume that both thermal buoyancy and shear effects 
contribute to the instability.  One may question whether it makes sense to do a linear
stability analysis in a solar model, since turbulent motions in the convective region imply that the stationary 
background necessary to derive a dispersion relation is not present 
\citep{M11}. While granulation and supergranulation flows on the solar surface and slightly below it indeed move vigorously at speeds $\sim 1$ \kms\, scaling arguments \citep{r10} suggest that supergranulation flows may be very shallow, confined to radii $r>0.99 ~\rr$. Recent helioseismic results indicate that large--scale turbulent velocities at $r\le0.96~\rr$  are typically quite small, of order 0.01 ~\kms\ \citep{hanasoge_seismic_2010, hanasoge_anomalously_2012}, while a similar helioseismic study of flows closer to the solar surface indicate that large--scale flows lose coherence below $ 0.99~\rr$  \citep{sv13}, while velocities drop significantly from the surface to $r \sim 0.99~\rr$.  The results of these studies suggest that application of a linear analysis is possible except very close to the solar surface.  With these assumptions, we find that there are indeed parts of the convective region of the Sun where the growth rate of MRI modes is more rapid than the 
growth of convective modes due to thermal buoyancy.

The paper is organized as follows.  In section \ref{sec:naximri}, we derive a dispersion relation for the triply diffusive nonaxisymmetric instability including the MRI and discuss the 
relevant instability criteria for that dispersion relation in the Sun. Section \ref{sec:methodology} describes our methodology for calculating the growth rate of modes in the Sun.   
Section \ref{sec:results} presents our results.  Section \ref{sec:discussion} compares our 
findings to previous work on the MRI in the Sun, and discusses the nonlinear saturation of shear modes including the MRI and convection in the Sun. Finally,
Section \ref{sec:conclusions} reviews our main conclusions.

\section{The Nonaxisymmetric MRI} \label{sec:naximri}

\subsection{Dispersion Relation} \label{sec:disprel}

We now calculate the growth rates of nonaxisymmetric, diffusive modes including the MRI that may be present in the Sun. To do this, we carry out a Wentzel$--$Kramers$--$Brillouin (WKB) perturbation analysis of the equations of magnetohydrodynamics (MHD) under the assumption that pressure 
perturbations are negligible except in the momentum equation
where they are coupled to thermal buoyancy effects (i.e., the Boussinesq 
approximation). We neglect composition gradients, because even the 
outer portions of the radiative region in the Sun are expected to have 
a homogeneous composition, and mixing in the convective region guarantees 
this homogeneity. The MHD equations under these approximations are
\begin{equation}
{\mathbf\nabla} \cdot \mathbf{v}  = 0,
\label{eq:mass}
\end{equation}

\begin{align}
\left(\frac{\partial }{\partial t} -\nu \nabla^2\right) {\mathbf v}+
  ( {\mathbf v\cdot \nabla} ) {\mathbf v} =& - \frac{1}{\rho }\nabla 
    \left( P+\frac{{\mathbf b}^2}{8\pi } \right) \nonumber \\&+ 
      \frac{1}{4 \pi \rho}( {\mathbf b\cdot\nabla} ) {\mathbf b} + {\mathbf g},
	\label{eq:momentum}
\end{align}

\begin{equation}\left(\frac{\partial }{\partial t} -\eta \nabla^2\right){\mathbf b} =
   ({\mathbf b\cdot\nabla} ) {\mathbf v}-({\mathbf v\cdot\nabla} ) 
     {\mathbf b}.
     \label{eq:induction}
\end{equation}

\begin{equation}
\left(\frac{\partial }{\partial t}+{\mathbf v \cdot \nabla}\right) \ln\frac{P}{\rho^\gamma} =\xi \nabla^2 \tau,
\label{eq:entropy}
 \end{equation}

In these equations, $\mathbf{b}$ is the magnetic field, $\mathbf{v}$ is the fluid velocity,  $\nu$ is the kinematic 
viscosity, $\eta$ is the magnetic resistivity, $\xi$ is the thermal 
diffusivity, $\mathbf{g}$ is the acceleration due to gravity, and 
$\tau\equiv T/T_0$ is a dimensionless temperature parameter normalized to 
the local equilibrium temperature $T_0$. Note that in Equation 
(\ref{eq:entropy}), we have assumed that the fluid may be treated as 
an ideal gas with adiabatic index $\gamma$. To complete our equation set, 
we may relate $\mathbf{g}$ to known thermodynamic 
quantities by assuming the initial mass distribution is in hydrostatic equilibrium. Because the 
equilibrium gravitational force is much larger than the equilibrium 
magnetic and shear forces, we may express $\mathbf{g}$ in cylindrical 
coordinates $(\varpi,\phi,Z)$ as: 
\begin{equation}
{\mathbf g}= \left(-\frac{1}{\rho}\frac{dP}{d\varpi}   ,0, -\frac{1}{\rho}\frac{dP}{dZ} \right).
\end{equation}

We now perform a local WKB analysis in cylindrical coordinates $(\varpi,\phi,Z)$, assuming that the perturbations are 
of the form $\delta \propto \exp \{i (k_\varpi \varpi + m\phi 
+ k_Z Z - \sigma t)\}$, where $m$ is an integer. We express oscillation 
frequencies in terms of $\omega \equiv \sigma - m \Omega$, which is the 
relevant oscillation frequency for disturbances in the rotating frame.  
Accounting for the effects of dissipation on these oscillations, we then 
introduce the variables 
\begin{equation} 
\omega_\alpha = \omega + i\alpha k^2  \ \ (\alpha = \xi , \eta , \nu).\label{eq:wdiss} 
\end{equation}

In order to use this WKB form for the perturbations, we must make the 
local approximation $m/\varpi \ll k_{\varpi}, k_Z$. As a result, the 
perturbation in the total pressure is negligible in the $\phi$ component 
of the momentum equation, and the nonaxisymmetric component of the 
continuity equation is negligible. As noted above, we also apply the 
Boussinesq approximation by setting $\delta P=0$ in all equations but 
the momentum equation, Equation (\ref{eq:momentum}).  We assume the local equilibrium magnetic 
field is uniform for simplicity; for weak initial fields, gradients in these fields are unlikely to have strong effects. Thus, we neglect magnetic buoyancy modes.   Finally, we make the assumption that the equilibrium magnetic field is primarily toroidal; 
i.e., $B_\phi \gg B_\varpi, B_Z$. Therefore, although we neglect 
$B_\varpi/\varpi$ relative to $k_\varpi B_\varpi$ due to the local 
approximation, we do \textit{not} neglect $B_\phi/\varpi$ and
$m B_\phi/\varpi$ relative to $k_\varpi B_\varpi$ in calculating the 
linearized equations. 

Keeping only linear order terms in equations 
(\ref{eq:mass})--(\ref{eq:entropy}), we find the following 8 equations 
for the 8 perturbed quantities (the three components of $\delta \mathbf{b}$ 
and $\delta \mathbf{v}$, $\delta P$, and $\delta  \rho$): 
\begin{equation}
k_\varpi\delta v_\varpi + k_Z\delta v_Z = 0, 
\label{eq6} 
\end{equation}
\begin{align}
i\omega_{\nu}\delta v_\varpi + 2\Omega\delta v_{\phi}=& i k_\varpi 
   \left(\frac{\delta P}{\rho} +\frac{\mathbf{B} \cdot \delta 
     \mathbf{b}}{4 \pi \rho}\right) 
      -\frac{i (\mathbf{k} 
        \cdot \mathbf{B})}{4 \pi \rho }\delta b_\varpi \nonumber \\&+\frac{2}{\varpi} 
          \frac{ B_{\phi}\delta b_{\phi}}{4 \pi \rho}-
             \frac{\delta\rho}{\rho^2}\frac{dP}{d\varpi}, 
\label{eq7} 
\end{align}
\begin{align}
i\omega_{\nu}\delta v_\phi - \frac {\kappa^2}{2\Omega}\delta 
  v_{\varpi}-\varpi\frac{d \Omega}{dZ}\delta v_{Z}=& 
    -\frac{i (\mathbf{k} \cdot \mathbf{B})}{4 \pi \rho }\delta b_\phi \nonumber \\ 
       &-\frac{1}{\varpi} \frac{ B_{\phi}\delta b_{\varpi}}{4 \pi \rho},
\label{eq8} 
\end{align}
\begin{align}
i\omega_{\nu}\delta v_Z =& i k_Z \left(\frac{\delta P}{\rho} +\frac{\mathbf{B} 
  \cdot \delta \mathbf{B}}{4 \pi \rho}\right) \nonumber \\ 
  &-\frac{i (\mathbf{k}   \cdot \mathbf{B})}{4 \pi \rho }\delta b_Z -\frac{\delta\rho}{\rho^2}\frac{dP}{dZ}, 
\label{eq9} 
\end{align}
\begin{equation}
\omega_{\eta}\delta b_\varpi = -(\mathbf{k} \cdot \mathbf{B}) \delta v_\varpi \;,
\label{eq10} 
\end{equation}
\begin{align}
i\omega_{\eta}\delta b_{\phi} + \displaystyle{\frac{B_{\phi}}{\varpi}}\delta v_\varpi 
  =&-\frac{d  \Omega}{d\ln \varpi}\delta v_{\varpi} \nonumber \\
  &-\varpi\frac{d \Omega}{dz}  \delta v_{Z}-i (\mathbf{k} \cdot \mathbf{B} )\delta v_{\phi}, 
\label{eq11} 
\end{align}
\begin{equation}
\omega_{\eta}\delta b_Z = -(\mathbf{k} \cdot \mathbf{B})\delta v_Z,
\label{eq12}
\end{equation}
\begin{equation}
i \omega_{\xi} \gamma\frac{\delta \rho}{\rho}+\delta v_\varpi 
     \frac{d \ln P\rho^{-\gamma}}{d \varpi} +\delta v_Z 
       \frac{d \ln P\rho^{-\gamma}}{d Z} =0,
\label{eq13}
\end{equation}
where we have defined the epicyclic frequency to be: 
\begin{equation} 
\kappa^2\equiv\frac{1}{\varpi^3}\frac{d \Omega^2 \varpi^4}{d\varpi} =4 \Omega^2 +\frac{d \Omega^2}{d \ln \varpi}
 \end{equation}
Note that in Equation (\ref{eq13}), we have eliminated the temperature 
perturbation using the relation $\delta \rho /\rho=-\delta T/ T$, which 
may be derived by combining the Boussinesq approximation $\delta P =0$ 
with the ideal gas law $P=\rho k_B T$.

Combining equations (\ref{eq6})-(\ref{eq13}), we find the triply--diffusive nonaxisymmetric dispersion 
relation including the MRI:
\begin{align}\label{eq:naxidisp}
\frac{k_{\rm pol}^2}{k_{Z}^2} {\widetilde\omega}_{\eta\nu}^4 -  {\widetilde N}^2\frac{\omega_\eta}{\omega_\xi}{\widetilde\omega}_{\eta\nu}^2  -{\widetilde \kappa}^2{\widetilde \omega}_{\eta}^2 -2{\omega_{\rm {A}\phi}}^2{\widetilde \omega}_{\eta \nu}^2\nonumber &\\ -4( \mathbf{k\cdot v_{\rm A}} )^2 \Omega^2 \left(1+\left[\frac{\omega_{\rm {A}\phi}} {\Omega}\right]^2 \right) \nonumber &\\ 
       -\frac{(\mathbf{k\cdot v_{\rm A}} )\omega_{\rm {A}\phi}}{\Omega} (\omega_{\eta}[4\Omega^2+ {\widetilde \kappa}^2 ]+\omega_{\nu}[4\Omega^2 -{\widetilde \kappa}^2 ]) &= 0 \;, 
\end{align}
where
\begin{equation}
k_{\rm pol}^2=k_\varpi^2+k_Z^2,
\end{equation}
\begin{align}
 {\widetilde\omega}^2_{\eta\nu} \equiv \omega_\eta\omega_{\nu} -(\mathbf{k\cdot v_{\rm A}} )^2 \;, \label{eq:wetanu}\\
 {\widetilde \omega}^2_\eta \equiv \omega_\eta^2 -(\mathbf{k\cdot v_{\rm A}})^2  \label{eq:wetasq}\;,  
\end{align}

\begin{equation}
\omega_{\rm {A}\phi}\equiv\frac{(\mathbf{k\cdot v_{\rm A}})_\phi}{m}=
  \frac{B_{\phi}}{\varpi\sqrt{4\pi \rho}},
\end{equation}
\begin{equation}
\label{Ntilde}
{\widetilde N}^2\equiv-\frac{1}{\rho \gamma}({\cal D} P)\, {\cal D} \ln (P\rho^{-\gamma}),
\end{equation}
\begin{equation}
{\widetilde \kappa}^2\equiv -\frac{1}{\varpi^3}\, {\cal D} (\varpi^4\Omega^2)=\kappa^2-\frac{k_\varpi}{k_Z}\varpi \frac{d\Omega^2}{dZ} ,
\end{equation}
	\begin{equation}
\qquad {\cal D} \equiv \left( \frac{k_{\varpi}}{k_Z}\frac{d}{d Z}-
   \frac{d}{d\varpi }\right).
	\end{equation}
	
The dispersion relation (\ref{eq:naxidisp}) is very similar to that found 
by \citet[their Equation 30]{MSS07}; the equations differ only because 
they neglect the shear in the $Z$ direction, which is important in the 
Sun, while we neglect neutrino radiation, which has a negligible effect 
in the Sun. If we neglect nonaxisymmetric effects completely by setting 
$\omega_{\rm {A}\phi}=0$, we recover the dispersion relation of 
\citet[their Equation 13]{MBS04}. Because this dispersion relation includes the full effects of shear and thermal buoyancy, it implicitly includes all axisymmetric modes of the Kelvin--Helmholtz and baroclinic instabilities. In Section \ref{sec:otherinstabilities} we discuss situations in the Sun where these instabilities may be active.

For modes on very large length scales, which correspond to small $k$, this dispersion relation can be simplified further.  For large--scale modes, the characteristic dissipative frequencies $k^2\xi$, $k^2\eta$, and $k^2 \nu$ are small compared to the rotation rate $\Omega$ and the magnitude of the buoyancy frequency $|N|$. Therefore, for all fast--growing modes with $|\omega|\sim \Omega$ or $|\omega|\sim |N|$, from Equation {\ref{eq:wdiss}} we have $\omega_{\xi}\sim  \omega_{\eta}\sim \omega_{\nu}\sim \omega$ and from Equations (\ref{eq:wetanu}) and (\ref{eq:wetasq}) we have ${\widetilde \omega_{\eta\nu}}^2\sim {\widetilde \omega_{\eta}}^2\sim {\widetilde\omega}^2$, where we define ${\widetilde\omega}^2 \equiv \omega^2 -(\mathbf{k\cdot v_{\rm A}} )^2$. The resulting dispersion relation is then
\begin{align}
\frac{k_{\rm pol}^2}{k_{Z}^2} {\widetilde\omega}^4 -  ({\widetilde N}^2+{\widetilde \kappa}^2+2{\waphi}^2){\widetilde \omega}^2\nonumber &\\
 -4( \mathbf{\kdotva} )^2 \Omega^2 \left( 1+2\frac{\waphi\omega} {(\kdotva) \Omega}+\left[\frac{\waphi} {\Omega}\right]^2\right)  &= 0. 
\end{align}

 For weak initial fields such that  $\waphi \ll \Omega $ and $\waphi\ll|N|$, small $k$ also implies that the characteristic magnetic frequency $\kdotva\ll \Omega$ and $\kdotva \ll |N|$. Therefore, for fast--growing modes with $|\omega|\sim \Omega$ or $|\omega|\sim |N|$, all terms that involve $\kdotva$ and $\waphi$ are negligible, and the dispersion relation becomes
\begin{equation}
\frac{k_{\rm pol}^2}{k_Z^2}  {\omega}^2-  {\widetilde N}^2   -{\widetilde \kappa}^2=0 \;. \label{eq:largescalemodes}
\end{equation} 

This final dispersion relation implies that for weak magnetic fields, large--scale modes are both adiabatic and hydrodynamic, since there is no coupling to the magnetic field or dissipation. 
  
In stars like the Sun, the thermodynamic variables density $\rho$ and pressure $P$, as well as the buoyancy frequency $N$, are typically functions only of spherical radius. In these cases, ${\widetilde N}$ from Equation (21) may be expressed in the form 
\begin{equation}
 {\widetilde N}^2=\left( \frac{k_\varpi}{k_Z}\cos \theta-\sin \theta\right)^2 N^2, \label{eq:ntildesph}
 \end{equation}
where 
	\begin{equation}
N^2=-\frac{1}{\rho \gamma}\frac{d P}{d r}\frac{d \ln P\rho^{-\gamma}}{d r}.\label{eq:nsqsph}
	\end{equation}
 is the square of the buoyancy frequency $N$, which is a function only of spherical radius.

We may also express ${\widetilde \kappa}$ in a simpler form as: 
 \begin{equation}
{\widetilde \kappa}^2=\kappa^2 -\frac{k_\varpi}{k_Z}\varpi \frac{d\Omega^2}{dZ}.
 \label{eq:kappatilde}
 \end{equation}
In the equatorial plane of such a star, ${\widetilde \kappa}=\kappa$ and ${\widetilde N}=N$.

\subsection{Instability Criteria} \label{sec:instcriteria}
We now calculate the instability criteria for this dispersion relation that are of importance in the Sun. 
Because we take the initial magnetic field to be weak, we can make 
the approximation  $\Omega\gg \omega_{\rm {A}\phi}$; we also eliminate 
the ``kink-type'' modes discussed by \citet{MST06} by focusing on modes 
in which $\mathbf{k\cdot v_{\rm A}}\gg \omega_{\rm {A}\phi}$.  Under 
this approximation, the manifestly nonaxisymmetric terms disappear, and the 
dispersion relation becomes  
\begin{equation}
\frac{k_{\rm pol}^2}{k_{Z}^2} {\widetilde\omega}_{\eta\nu}^4 -  {\widetilde N}^2\frac{\omega_\eta}{\omega_\xi}{\widetilde\omega}_{\eta\nu}^2   -{\widetilde \kappa}^2{\widetilde \omega}_{\eta}^2 -  4( \mathbf{k\cdot v_{\rm A}} )^2 \Omega^2 =0.\label{eq:axidisp}
\end{equation}

 It is important to note that because a toroidal field is present,  $\mathbf{k\cdot v_{\rm A}}$, and therefore the dispersion relation, still has a dependence on the nonaxisymmetric wavenumber $m$. This dispersion relation, Equation (\ref{eq:axidisp}), is identical in form to that of \citet{MBS04} for the axisymmetric instability. In what follows, we will make use of their results in the case of spherically symmetric contours of density $\rho$ and pressure $P$ to calculate the instability criteria for the dispersion relation in various limits that are relevant in the Sun. Because the ordering of the diffusion parameters in the Sun is $\xi \ggg \eta \gg\nu$ (see Section \ref{sec:methodology}), the appropriate conditions for stability are those for the limit $\nu \rightarrow 0$, given by \citet{MBS04} Equations 21, 37, 50, 56, and 62. Written in our notation, these conditions are

\begin{eqnarray}
{\widetilde N}^2 +{\widetilde \kappa}^2  >0, \label{eq:mricondfirst}\\
 {\widetilde N}^2 +{\widetilde \kappa}^2 -4\Omega^2 >0, \\
 2\frac{\eta}{\xi}{\widetilde N}^2 + (1+\frac{\eta}{\xi}){\widetilde \kappa}^2 >0, \\
 \frac{\eta}{\xi}{\widetilde N}^2 +{\widetilde \kappa}^2 -4\Omega^2 >0, \\
 {\widetilde N}^2 >0. 
 \label{eq:mricondlast}
\end{eqnarray}

 We will now discuss which of these stability conditions are violated in various locations in the Sun. Equation (\ref{eq:ntildesph}) implies that ${\widetilde N}^2$ has the same sign as $N^2$; therefore, there will be major differences between the stability characteristics of the dispersion relation in stably stratified regions and in convectively unstable regions, and we will treat them separately. 
 
\subsubsection{Stably Stratified Regions}\label{sec:stratcriteria}
In the radiative zone and the lower tachocline, the Sun is strongly stratified, with $N^2\gg \Omega^2 >0$.  Equation (\ref{eq:ntildesph}) then implies that ${\widetilde N}^2$ is a positive definite quantity in these regions.  Then, there are two possible ways in which the conditions \ref{eq:mricondfirst}--\ref{eq:mricondlast} may be violated. Firstly, there are a small set of modes for which the wavevector is very close to being in the $\pm\theta$ direction; this corresponds to $\left| k_\varpi/k_Z+\tan\theta\right|\ltsim\Omega/N$. These modes are typically unimportant in the Sun.
 
 The other case corresponds to ${\widetilde N}^2 \gg \Omega^2 \sim {\widetilde \kappa}^2$. In this case, we may neglect factors of $\eta/\xi$ that are not multiplied by ${\widetilde N}^2$. Comparing the five stability conditions then indicates the necessary and sufficient stability criterion is  
\begin{equation}
\frac{\eta}{\xi}{\widetilde N}^2 +{\widetilde \kappa}^2 -4\Omega^2 >0,
 \label{eq:mridisscond}
\end{equation}

Following \citet{MBS04}, this criterion implies that unstable modes exist if

\begin{equation}
 \frac{\eta}{\xi}N^2 +\frac{d \Omega^2}{d \ln \varpi}<0,
 \label{eq:mridisscondr}
\end{equation}
or if
\begin{equation}
 \left(\varpi \frac{d \Omega^2}{d Z}\right)^2-8\frac{\eta}{\xi}N^2 \sin(\theta)\cos(\theta)\frac{d\Omega^2}{d\theta}>0. 
\label{eq:mridisscondthz}
\end{equation}

 Note that we again neglect factors of $\eta/\xi$ that are not multiplied by ${\widetilde N}^2$. The first condition represents the destabilizing influence of cylindrically radial shear, which is opposed by stable stratification; while the second represents the destabilizing influence of shear in the $Z$ and $\theta$ directions.  Both of these conditions for instability correspond to small-scale magnetohydrodynamic modes driven by shear, which can properly be called MRI modes.

\subsubsection{Convectively Unstable Locations}
In convectively unstable regions with $N^2<0$,  Equation (\ref{eq:ntildesph}) implies that ${\widetilde N}^2$ is also negative. Therefore, thermal buoyancy effects always contribute to instability, and modes can be driven by the combined effects of convection and shear. The instability criteria for these modes are generally quite complicated. Any or all of the conditions \ref{eq:mricondfirst}--\ref{eq:mricondlast} may be violated; however, two limiting cases exist in which the instability criteria are more tractable.  In the limiting case where $\eta/\xi \left|N^2\right|\gg \Omega^2$, rotational effects are negligible. Then, all of the stability conditions reduce to

\begin{equation}
{\widetilde N}^2>0.
 \label{eq:mriallconv}
\end{equation}

The resulting conditions where instability can occur are 

\begin{equation}
N^2 <0,
\end{equation}
and
\begin{equation}
\frac{d \Omega^2}{d \theta}>0.
\end{equation}

Because the first inequality is always satisfied in the convective zone, it is generally the important one for this case; it corresponds directly to hydrodynamic convective modes on large scales.  

The other limiting case occurs when convective effects are negligible, which corresponds to $\Omega^2 \gg \left|N^2 \right|$. Then, the sufficient instability criteria take the form:
\begin{equation}
{\widetilde \kappa}^2 -4\Omega^2 >0.
\end{equation}

The conditions under which instability can occur are given by

\begin{equation}
\frac{d \Omega^2}{d \ln \varpi}<0,
\end{equation}
and
\begin{equation}
\frac{d \Omega^2}{d Z} \neq 0. \label{eq:shearzmodes}
\end{equation}
The first inequality corresponds to small--scale MRI modes that are similar to those to those found in the stably stratified regions. The second inequality is always violated unless rotation is constant on cylinders; it corresponds to large--scale hydrodynamic shear modes. 

\subsubsection{The Kelvin--Helmholtz and Baroclinic Instabilities}
\label{sec:otherinstabilities}
 We now discuss where the Kelvin--Helmholtz and baroclinic instabilities may be active in the Sun.    The condition for the Kelvin--Helmholtz instability for a plane--parallel geometry in a fluid with continuously varying density and velocity and no interfaces is $Ri<1/4$, where the Richardson number, $Ri$, is defined as  $Ri=N^2/(dv/dy)^2$, where $v=\Omega r$ and $y$ is both the direction of gravity and the direction in which the velocity of the fluid varies. In a star with predominantly radial shear, these directions coincide at the equator, so we make the identification $y=r$. Then, the instability criterion may be written as

\begin{equation}
N^2-(1/4)\Omega^2 (1+q)^2<0
\end{equation}

\noindent where $q=d\ln \Omega/d \ln r$.  In the stably stratified tachocline, this instability criterion will only be satisfied in a very narrow region at $r\approx0.713\rr$, because $N^2\gg\Omega^2$ at all locations for which $N^2>0$. In the convective zone, the instability is always active because $N^2<0$, but it will be significantly modified by the presence of convective instability except in the convectively unstable tachocline, where $|N^2|\ll \Omega^2$ and $q\sim 1$. In these locations, it may have similar effects to the hydrodynamic shear modes discussed later in this paper.

The baroclinic instability occurs in rapidly rotating, stably stratified
environments when the gradients in pressure and density are not aligned.
The baroclinic instability arises in conditions of small Rossby number,
$Ro = v/Lf$, the ratio of inertial to Coriolis force terms, where $v$ is a 
characteristic velocity, $L$ a characteristic length scale, and $f = 2 \Omega \cos \theta$ is the Coriolis frequency. A small Rossby number indicates conditions strongly 
affected by Coriolis forces; a large Rossby number indicates that inertial and 
centrifugal forces dominate. Because the baroclinic instability involves vortical flow, it is difficult to characterize in our shellular calculations, in which thermodynamic variables are assumed to be spherically symmetric.  But in general, the baroclinic instability and the Kelvin--Helmholtz instability are likely to be active in similar regions. Strong stratification, indicated by a large Richardson number, is likely to inhibit the growth of the baroclinic instability in the radiative region. It may be active to some extent
in the tachocline, but is probably overwhelmed by other influences in the
convective regions of the Sun.

\section{Methodology}\label{sec:methodology}

We now calculate the growth rates of unstable modes of the triply--diffusive nonaxisymmetric dispersion relation throughout the Sun. Table \ref{tab:parameters} presents the important parameters and variables used in this calculation. To calculate thermodynamic 
variables, we compute a 1D model of the Sun using MESA \citep{paxton11}; 
we thus assume that all thermodynamic variables, such as $P$ and $\rho$, 
are functions only of spherical radius $r$. We note that the one dimensional non--rotating solar model we have computed is not self-consistent with the known rotational profile of the Sun or the initially assumed magnetic fields; however, adoption of this Solar model
is a necessary first step to a deeper understanding of the rotating, magnetic
evolution of the Sun and other stars.

 \begin{deluxetable}{ll}
  \scriptsize
  \tablecolumns{2}
  \tablecaption{Coordinates and parameters in this paper}
\tablehead{\colhead{Parameter} & \colhead{Description}}
\startdata 

$\varpi$ & Cylindrical radius \\
$Z$ & Height above equatorial plane\\
$\phi$ & Azimuthal angle \\
$\theta$ &Spherical colatitude\\
$r$& Spherical radius\\

$\mathbf{B}=(B_\varpi,B_\phi, B_Z)$ &Initial magnetic field\\
$B_{\rm pol}=\sqrt{B_\varpi^2+B_Z^2}$ & Initial poloidal magnetic field\\
$R_{\rm TP}=  B_\phi/ B_{\rm pol}$  & Toroidal to poloidal field ratio \\
$\mathbf{k}=(k_\varpi,k_\phi, k_Z)$ &Instability wavenumber\\
$m=k_\phi \varpi$ &Dimensionless toroidal wavenumber \\
$B_\varpi/B_Z$& Direction of poloidal magnetic field\tablenotemark{a} \\
$k_\varpi/k_Z$& Direction of poloidal wavenumber\tablenotemark{a}\\
$\delta v_\varpi/\delta v_Z$& Direction of poloidal displacement\tablenotemark{a}\tablenotemark{b} \\
$\Phi_{JK}$& The angle between poloidal vectors $\mathbf{J}$ and $\mathbf{K}$\\

$v_{\rm A}=B/\sqrt{4\pi \rho}$&Alfv\'en velocity\\
$\rho$  &Density\\
$P$&Pressure\\
$\xi$  &Thermal diffusivity\\
$\eta$ &Magnetic resistivity\\
$\nu$ &Kinematic viscosity\\
$\Omega$ &Angular velocity of the star \\
$N$ & Brunt--V\"{a}is\"{a}l\"{a} (Buoyancy) frequency\\
$\Gamma$& Growth rate of instability\\
$\kdotva$ & Coupling between wavenumber and magnetic field\\
$\waphi=B_{\phi}/\varpi\sqrt{4\pi \rho}$ & Initial toroidal field strength in frequency units \\
$k^2 \xi$ & Thermal dissipation frequency \\
$k^2\eta$ &Resistive dissipation frequency\\
$k^2\nu$ &Viscous dissipation frequency\\
$\mu=m\waphi/\Omega$ & Normalized nonaxisymmetric contribution to $\kdotva$\\
$q=d \ln \Omega / d \ln r$ & Dimensionless spherically radial shear\\
Region TS &Stably stratified tachocline \\
Region TU\tablenotemark{c} & Lower convective zone ($\theta<60 \degrees$)\\
Region TL\tablenotemark{c} & Lower convective zone ($\theta>60 \degrees$)\\
Region C & Upper convective zone
\enddata
\label{tab:parameters}
\tablenotetext{a}{The physical meaning of these quantities is explained in \\more detail in Section \ref{sec:methodology}}
\tablenotetext{b}{Equation (\ref{eq6}) shows that $\delta v_\varpi/\delta v_Z=-(k_\varpi/k_Z)^{-1}$}
\tablenotetext{c}{These regions also include the convectively unstable portion of the tachocline at the given values of $\theta$}
\end{deluxetable} 
 Following \citet{MBS04}, we calculate the values of the three diffusivities in the Sun from the thermodynamic quantities in the MESA model. The thermal diffusivity is dominated by radiative transport, and is given by  
\begin{equation}\xi=\frac{\gamma-1 }{\gamma}\frac{T}{P}\frac{16T^3}{3\kappa \rho },\end{equation}
 where $\kappa$ is the radiative opacity. 
 
The resistivity $\eta$ is given by 
\begin{equation}
\eta\approx 5.2 \times 10^{11}\frac{\ln \Lambda}{T^{3/2}} \, {\rm cm^2}\, {\rm s^{-1}},
\end{equation} 
where $\ln \Lambda$ is the Coulomb logarithm. This logarithm is given in the NRL plasma formulary as
\begin{equation}
\ln \Lambda \approx \begin{cases} -17.4 +1.5\ln T -0.5 \ln \rho & \,   T < 1.1\times10^5 \, \rm{K},\\
        -12.7+ \ln T -0.5 \ln \rho &\,   T > 1.1\times10^5 \, \rm{K}.
\end{cases}
\end{equation}
after translating into cgs units.

The viscosity $\nu$ is dominated by thermal viscosity, and is given by
\citet{spitzer06} as
\begin{equation}
\nu\approx 5.2 \times 10^{-15}\frac{T^{5/2}}{\rho \ln \Lambda} \,   {\rm cm^2}\, {\rm s^{-1}}.
\end{equation}

We calculate the gradients of $\Omega$, in the convective and radiative zones 
using GONG helioseismic data \citep{howe09}. These data lack fine spatial resolution in the tachocline and therefore greatly underestimate the radial shear there. In the region of the tachocline, we instead estimate the radial shear using the approximate equation \citep{PM07}
\begin{equation} 
\frac{\partial \Omega}{\partial r}= \frac{\delta \Omega_{\rm eq}}{\Delta}\left( 1- 3.56\cos^2 \theta - \cos^4 \theta\right),
\end{equation}
where $\delta \Omega_{\rm eq}= 1.08\times10^{-7} $ rad/s is the change in angular velocity across the tachocline at the equator, and $\Delta=0.02~\rr$ is the width of the tachocline. 

We then solve Equation (\ref{eq:naxidisp}) for $\omega$ as a function 
of location in the Sun, the wavenumber $\mathbf{k}=(k_\varpi,m/\varpi,k_Z)$, and the initial magnetic field $\mathbf{B}=(B_\varpi,B_\phi,B_Z)$. The components of the magnetic field and the wavenumber appear in the dispersion relation (\ref{eq:naxidisp}) solely through their contributions to $k_\varpi/k_z$,$k^2$, $(\kdotva)^2$, $\waphi^2$, and $(\kdotva)\waphi$. For values of $m$ that obey the local approximation, the magnitude of the wavevector is $k\approx k_{\rm pol}$. Therefore, we can specify the complete parameter space by setting $k$, $B_{\rm pol}=\sqrt{B_\varpi^2+B_Z^2}$, $k_\varpi/k_z$,   $B_\varpi/B_z$, $m$, and $R_{\rm TP}\equiv B_{\phi}/B_{\rm pol}$, which is the ratio of the toroidal and poloidal magnetic field components. The physical significance of the ratios $k_\varpi/k_Z$ and $B_{\varpi}/B_Z$ is that they specify the direction of the poloidal wavenumber ${\mathbf k}_{\rm pol}$ and the poloidal magnetic field ${\mathbf B}_{\rm pol}$. Two other ratios that are not free parameters are the ratio $\varpi/Z=\tan \theta$, which specifies the direction of the spherical radius $\mathbf{r}$ at a given location, and $\delta v_\varpi/\delta v_Z$, which specifies the direction of the displacement resulting from the perturbed velocity $\delta { \mathbf v}$ and is related to $k_\varpi/k_Z$ by Equation (\ref{eq6}). 

The angle $\Phi_{JK}$ between the directions of two poloidal vectors $\mathbf{J}$ and $\mathbf{K}$ is related to the dot product of the vectors $\mathbf{J \cdot K}$ and may be expressed in terms of their magnitudes $J$ and $K$ and the ratios $J_\varpi/J_Z$ and $K_\varpi/K_Z$ using the equation
\begin{align}
\cos \Phi_{JK}=\frac{\mathbf{J\cdot K}}{JK}=&\left(1+\frac{J_\varpi}{J_Z}\frac{K_\varpi}{K_Z}\right) \nonumber \\
   \times & \frac{{\rm sgn}(J_Z K_Z)}{\sqrt{(1+(J_\varpi/J_Z)^2)(1+(K_\varpi/K_Z)^2)}}  \label{eq:dotproduct} ,
\end{align}
where ${\rm sgn}(J_Z K_Z)=J_Z K_Z/|J_Z K_Z|$ is the sign of the product of the $Z$-components of the vectors. In our axisymmetric calculations, we typically find that only the magnitude of $\cos \Phi_{JK}$ is of importance in a calculation and this sign is insignificant; we may then take $\Phi_{JK}$ to have domain $-90 \degrees<\Phi_{JK}\leq 90 \degrees$.

Equation (\ref{eq:dotproduct}) makes it possible to simply compare the directions of poloidal vectors specified by referring to their magnitudes and the ratio of their poloidal components. Equation (\ref{eq:dotproduct}) is also used to calculate the poloidal parts of the dot product $\kdotva=(\mathbf{k}\cdot \mathbf{B})/\sqrt{4 \pi \rho}$ that appears in the dispersion relation (\ref{eq:naxidisp}) from the specified parameters $k_{\rm pol}$ and $B_{\rm pol}$, $k_\varpi/k_Z$, and $B_{\varpi}/B_Z$.   

We now specify the ranges of each of the parameters that characterize the solutions of the dispersion relation; we first choose the appropriate ranges for the wavenumber $k$. The local approximation creates a constraint requiring that the mode wavelength $\lambda=2\pi/k$ fits within a single local pressure scale height $H_{\rm P}$; a lower limit on $\lambda$ is provided by the effects of magnetic resistivity and viscosity, which stabilize modes on scales for which $k^2 \eta$, $k^2 \nu \gg k v_{\rm A}$. The local approximation also requires that $\left|m \right|/\varpi\ll k$; we set $\left|m \right|\le   15$ to ensure that the local approximation is satisfied even for the largest-scale modes with $k=2\pi/H_{\rm P}$; the maximum value of $H_{\rm P}$ in the tachocline and convective region is always smaller than $0.1~\rr$. We investigate the growth rates of small--scale modes with larger toroidal wavenumbers in Section \ref{sec:naxieffects}. Because the final wavenumber parameter $k_\varpi/k_Z$ is not constrained by our approximations, we vary it freely, including both positive and negative values with magnitudes greater than or less than 1.

We next choose an initial magnetic field strength and geometry. We 
assume a small poloidal magnetic field of magnitude $0.2$ G and explore the effects of varying this magnitude in Section \ref{sec:fielddependence}. To explore the effects of the poloidal magnetic field geometry, we set $B_{\varpi}/B_Z$ corresponding to a field oriented in the $r$, $ \theta$, $ \varpi$, and $Z $ directions; reversing the direction of this field is equivalent to making the substitution $m \rightarrow -m$, so it is unnecessary to consider the opposite orientations. We determine the strength of the toroidal magnetic field at a given point by setting $R_{\rm TP}=5$, consistent with the expected dominance of the toroidal magnetic field in stellar 
MHD equilibria \citep{Braith09}; we investigate the effects of varying $R_{\rm TP}$ in Section \ref{sec:naxifieldgeometry}. Setting $R_{\rm TP}$ fixes the values of $\omega_{\rm {A}\phi}$ and $(\mathbf{k\cdot v}_{\rm A})_\phi$.  Having set the ranges of the parameters, we then
calculate the nonaxisymmetric growth rate $\Gamma=-i \omega$ for  
$\sim10^6$ wavevectors in the phase space, and compare the results for all of the indicated field geometries at each location in 
the Sun.

\section{Results}\label{sec:results}

We now discuss our calculations of the growth rates of modes throughout the Sun. In Section \ref{sec:mrigrowth}, we present the growth rates of the most rapidly growing modes of the overall instability in the Sun and discuss whether shear or convection is responsible for driving the instability at each location. In Section \ref{sec:aximodes} we discuss the nature of the axisymmetric instability throughout the Sun and identify those modes that are sensitive to the initial magnetic field geometry. Finally, in Section \ref{sec:naxieffects}, we discuss nonaxisymmetric effects and their variation with the toroidal field and the initial poloidal magnetic field strength.

\subsection{Growth Rates of Instability in the Sun} \label{sec:mrigrowth}

\begin{figure}[t]
\begin{center}
\includegraphics[width = 0.45\textwidth]{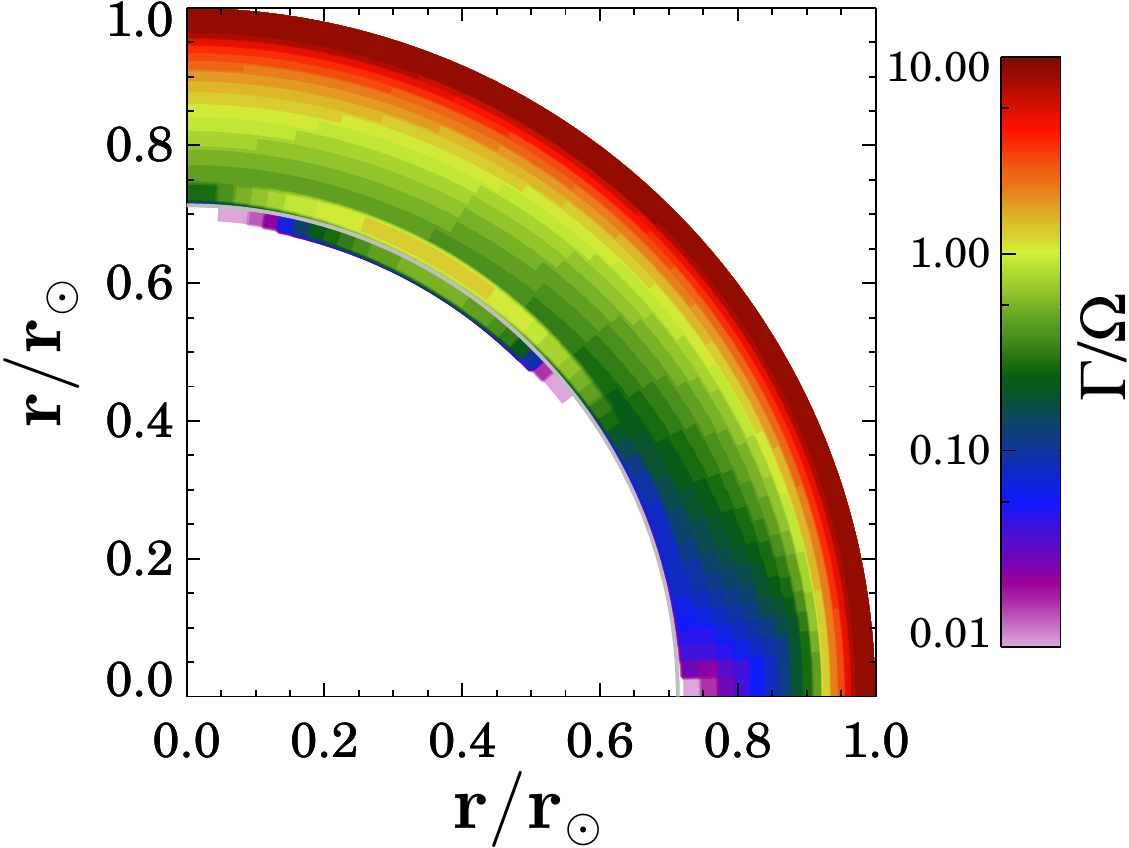}
\end{center}
\caption{{\small The growth rate $\Gamma$ of the instability throughout the Sun in terms of the local rotation rate $\Omega$. The gray line indicates the bottom of the convection zone in the tachocline, where $N=0$. For definiteness, we do not indicate that growing modes are present unless their growth rate is larger than $0.01\Omega$. The rapid growth of this instability throughout the convective zone indicates that it may play a role in the origin of the solar magnetic field.}\label{fig:mrigrowth}}
\end{figure}

\begin{figure}[t]
\begin{center}
\includegraphics[width = .45\textwidth]{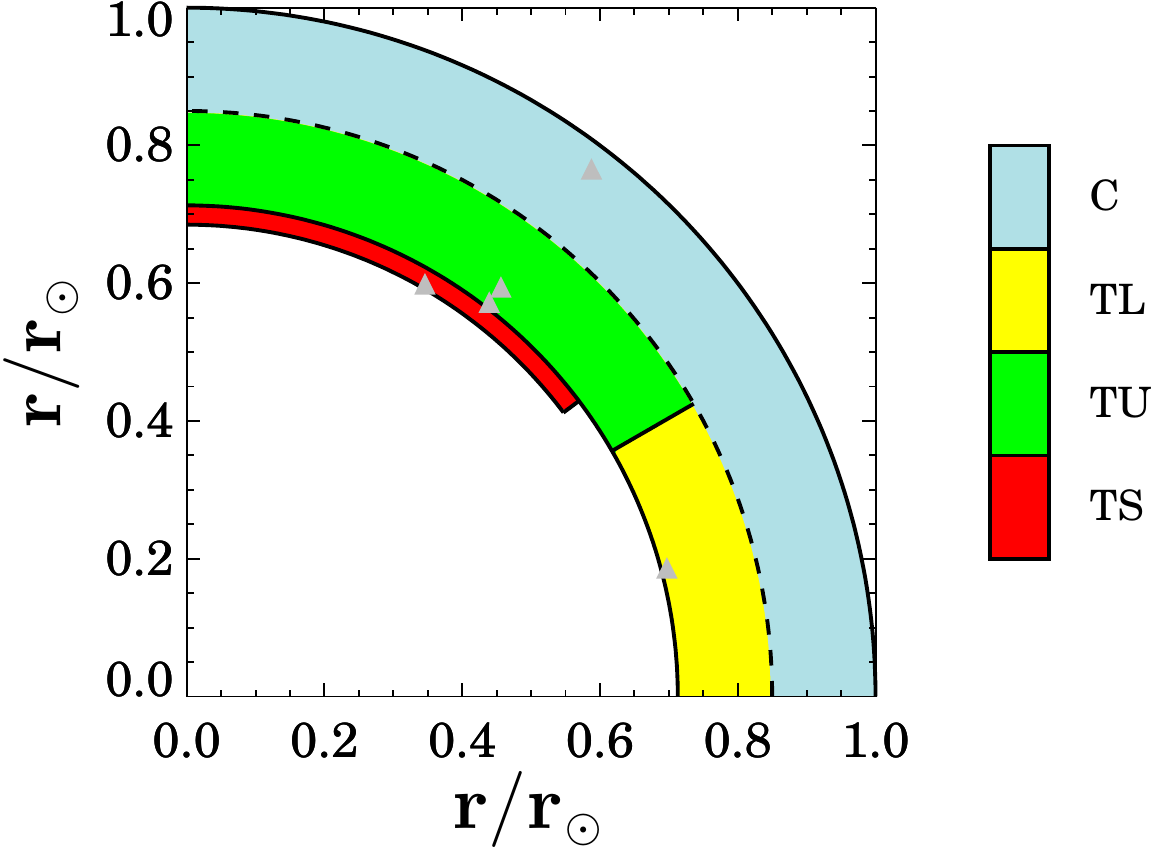}
\end{center}

\caption{{\small Locations of the four regions of the Sun that determine the characteristics of growing modes. The center of each gray triangle shows a point where detailed calculations are carried out in Section  \ref{sec:aximodes}. The properties of each region are described in this caption as:\\ Region TS, which has $N^2>0$, $q<0$, and $|N| \gg\Omega$\\  Region TU, which has $N^2<0$, $q<0$, and $|N| <\Omega$ \\Region TL, which has $N^2<0$, $q>0$, and $|N| <\Omega$\\ Region C, which has $N^2<0$, varying values for $q$, and $|N| \gg \Omega$.  \\ The dashed line indicates that there is no precise boundary between Region C and Regions TU and TL; instead, the strength of convection increases gradually with radius, becoming dominant at large radius (see text).  }\label{fig:regions}}
\end{figure}

 Figure \ref{fig:mrigrowth} shows the maximum growth rate, $\Gamma$, of the unstable modes of the dispersion relation Equation (\ref{eq:naxidisp}) at each location in units of the local angular rotation velocity $\Omega$; for this calculation, the maximum growth rate is calculated for any poloidal field geometry. It is clear that the instability grows quickly throughout the tachocline and the solar convective region. The Sun may be divided into four regions in which the characteristics of the most rapidly growing modes have significant differences; Figure \ref{fig:regions} presents the locations of these regions, the typical radial shear $q$, and qualitative comparisons of the characteristic frequencies $\Omega$ and $N$. In the subsequent discussion, we will refer to ``upper," ``lower," ``top," and ``bottom," (as in, e.g., the upper tachocline and the bottom of the convective region) in terms of the spherical radial coordinate. 

Region TS is located in the stably stratified part of the tachocline at $r<0.713~\rr$. The most rapidly growing modes in this region are nonaxisymmetric, with the maximum growth rate corresponding to the largest value for the toroidal wavenumber $|m|$; we discuss nonaxisymmetric effects in Section \ref{sec:naxieffects}. Region TU is located in the convectively unstable region close to the tachocline at $r>0.713~\rr$, has a colatitude range of $0\degrees<\theta < 60 \degrees$ that is similar to, but slightly larger than, that for Region TS, and corresponds closely to the region for which the radial shear in the tachocline is negative. 

In both Region TS and Region TU, the most rapidly growing modes have growth rate $\Gamma$ on the order of $\Omega$, although the growth rates tend to be significantly lower very close to the poles and for $\theta> 45 \degrees$. Note that the somewhat smaller growth rates in Region TS result from the fact that the GONG measurement of $\Omega$ in this region is taken at $r=0.692~\rr$, which is at the bottom of the tachocline; a calculation using a slightly smaller value for $N^2$ corresponding to $r\sim 0.70 ~\rr$ gives similar growth rates to those found in Region TU.  The growth rate in both Regions TS and TU is maximized at $\theta\sim20\degrees$--$30\degrees$, which is the approximate location in the tachocline where the shear in the $\varpi$ direction is largest. This indicates that the strong shear in the tachocline is probably driving the growth of instability. The lack of growth of modes in Region TS for colatitudes $53 \degrees<\theta <60 \degrees$ is probably a result of the strong stable stratification in this region, where the radial shear is relatively weak. 
 
  Region TL is located at lower latitudes corresponding to $\theta >60 \degrees$ in the convectively unstable region $r>0.713~\rr$ of the tachocline and lower convective zone.   The typical growth rate of instabilities in Region TL is typically much smaller than both the rotation rate $\Omega$ and the magnitude of the buoyancy frequency $\left| N \right|$; this is likely because the positive radial shear retards the growth of instability. Finally, in Region C, located at $r> 0.8~\rr $ at all latitudes, the typical growth rate of the most rapidly growing modes is similar to the local magnitude of the buoyancy frequency $\left| N \right|$. This indicates that the modes in this region are driven by convection. It should be noted that there is no precise physical boundary between Region C and Regions TU and TL; instead, the strength of convection increases gradually with radius, becoming dominant at large radius. 

\begin{figure}[t]
\begin{center}
\includegraphics[width = .45\textwidth]{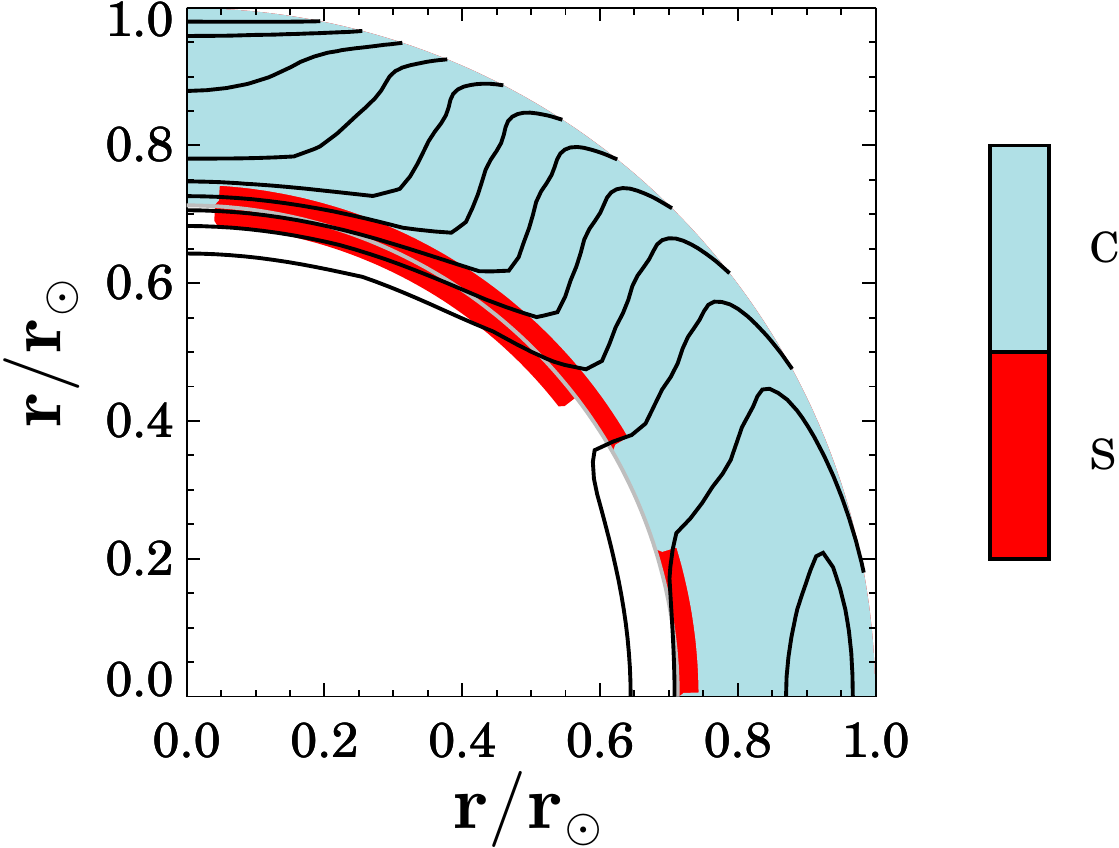}
\end{center}

\caption{{\small The locations in the Sun where the growth of instability is dominated by convection (shown in blue) and shear (shown in red). The black lines are contours of $\Omega$ found by GONG; they show the layers of strong shear in the tachocline and near the solar surface. Shear is dominant at most latitudes close to the tachocline, while convection dominates near the tachocline at $\theta \sim 60 \degrees$ and throughout the rest of the convective zone.  }\label{fig:convshear}}
\end{figure}

  In order to more precisely determine whether convection or shear is responsible for the growth of instability in each location, we repeat the calculation of the growth rate with only convection present (by setting all derivatives of the rotation rate $\Omega$ to 0) and with only shear present (by setting the buoyancy frequency $N=0$) and compare the resulting growth rates. Figure \ref{fig:convshear} shows those locations where the growth of instability is driven by convection and by shear. The figure indicates that shear is the only mechanism that can drive instability in the stably stratified Region TS, as expected from the analysis in Section \ref{sec:stratcriteria}. At the bottom of Region TU, shear is the dominant driver of instability, in agreement with our previous conclusions. The driver of modes at the bottom of Region TL at lower latitudes varies; modes are driven by convection at $60 \degrees<\theta <73 \degrees$, and by shear for $73 \degrees<\theta <90 \degrees$. We will discuss the cause of this variation in Section \ref{sec:phasetl}. In the rest of the convectively unstable region of the Sun, including the upper parts of Region TL and TU and the entirety of Region C, convection is responsible for driving the growth of instability. 

Figure \ref{fig:convshear} also shows contours of $\Omega$ obtained from GONG helioseismology, which are consistent with the description of the solar rotation profile discussed in Section \ref{sec:intro}. The region of strong shear near the tachocline drives unstable modes, but we show in Section \ref{sec:phasec} that the region close to the solar surface does not; we discuss the reason for this in detail in Section \ref{sec:mricomparison}. We use the quantitative ratio of the growth rates with only shear present and with only convection present to set the boundary between Regions TU/TL and Region C.  For $r>0.85~\rr$, the growth rate of modes driven by convection is at least 10 times higher than that of modes driven by shear at all values of $\theta$. We therefore set our fiducial boundary between Region C and Regions TU/TL at $r=0.85~\rr$.

\subsection{Axisymmetric Mode Analysis} \label{sec:aximodes}
In this section, we discuss the variation of the growth rate $\Gamma$ with initial parameters for axisymmetric modes with $m=0$.  We find that for axisymmetric modes, there is no significant dependence of growth rate on the sign of $\kdotva$; therefore, in our analysis of variation with wavenumber we present the variation of growth rate with $k_\varpi/k_Z$ and $|\kdotva|$. Our analysis of the variation of growth rate with the wavenumber is divided into four parts, corresponding to the four regions of the Sun in which growth of instability can occur discussed in the previous section: Regions TS, C, TL, and TU. We also discuss the submodes of the instability in each of these region, which may be driven by convection or shear; a full listing of what modes are present in each region is given in Table 2, which also summarizes our overall conclusions. We choose an initial magnetic field geometry with $|B_{\rm pol}|=0.2 G$, $B_{\varpi}/B_Z=\tan \theta$, which corresponds to a magnetic field oriented in the $r$ direction, and a toroidal to poloidal field ratio of $R_{\rm TP}=5$.
\subsubsection{Stably Stratified Region TS} \label{sec:phasets}

In the stably stratified tachocline corresponding to Region TS, only shear can drive unstable modes. While the radial shear $d \Omega^2/d \ln \varpi$ is negative for $\theta<62 \degrees$, we find modes with significant growth only for $\theta<53 \degrees$.
\begin{figure}[h]
\begin{center}
\includegraphics[width = .45\textwidth]{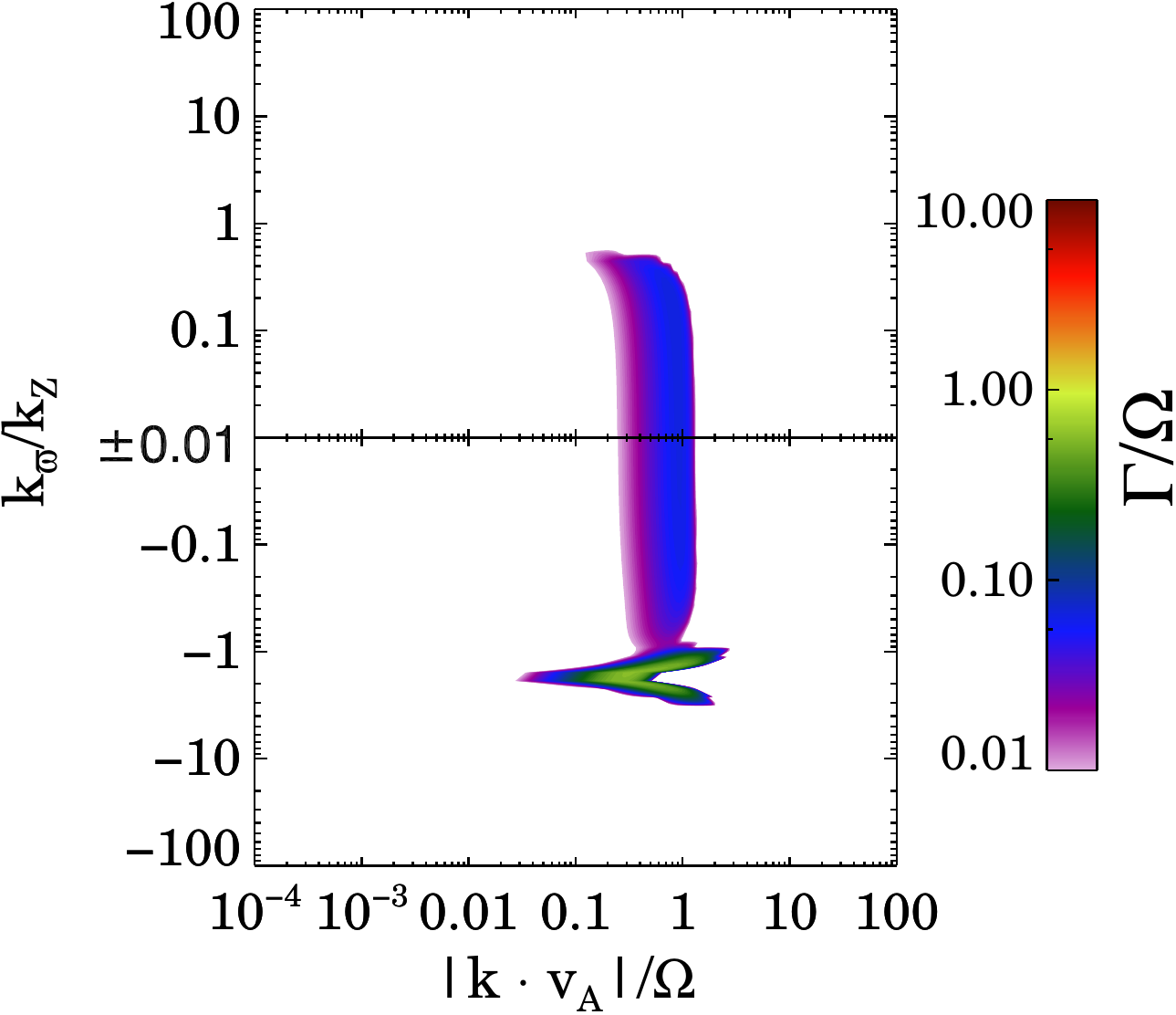}
\end{center}

\caption{{\small The dependence of the growth rate $\Gamma$ on the phase space parameters $k_\varpi/k_Z$ and $\left|\kdotva\right|$ at the location $r=0.692 ~\rr$, $\theta=30\degrees$ in Region TS (Figure 2); the growth rate and $|\kdotva|$ are given in units of the local rotation rate $\Omega$. The figure shows two regions of fast growth; the first is the CMRI mode  with $k_\varpi/k_Z\ll 1$ and $|\kdotva|\approx 0.5-1.0\Omega$, while the second is the PSS mode with $k_\varpi/k_Z\approx-1.5$ and a wide range of $|\kdotva|\approx$0.03$\Omega$--2.0$\Omega$. Note that the two wings of the PSS mode correspond to opposite signs of $(\kdotva)_{\rm pol}$.  The most rapidly growing mode has growth rate $\Gamma=0.63\Omega$, and is located in the PSS region at the coordinates $k_\varpi/k_Z=-1.57$, $|\kdotva|=0.32\Omega$.}\label{fig:phasets}}
\end{figure}

Figure \ref{fig:phasets} shows the variation with $k_\varpi/k_Z$ and  $|\kdotva|$ of the growth rate $\Gamma$ at $r=0.692~\rr$, $\theta=30 \degrees$ in Region TS. It clearly shows two major regions of instability: first, a region at low $|k_\varpi/k_Z|\ll 1$ and $|\kdotva|\approx0.5 \Omega$, and second, a region at $k_\varpi/k_Z \approx-1.57$ with a wider variation of $|\kdotva|$. The first region corresponds to nearly vertical modes, which do not couple to shear in the $Z$ direction or to the magnetic field in the $\varpi$ direction. These modes therefore couple to the star as though it were cylindrically symmetric, and are thus analogous to the classical MRI presented by \citet{BH91} in accretion disks; the only difference is that the shear profile is non--Keplerian and thermal buoyancy effects in the $\varpi$ direction are present. We therefore call this the classical MRI (CMRI) mode.  The most rapid growth rate in this region is $\Gamma=0.073\Omega$, and it is located at $k_\varpi/k_Z=0.16$, $|\kdotva|=0.83 \Omega$. 

The second region of phase space where fast growth occurs corresponds to modes with wavenumber nearly perpendicular to the magnetic field, $ \Phi_{Bk}\approx90 \degrees$, which corresponds to $|\kdotva|\ll k v_{\rm A}$; we call these modes perpendicular small scale (PSS) modes. PSS modes have large $k$, a condition that greatly weakens the stable stratification, but the moderate value for $\kdotva$ means that the magnetic tension does not greatly reduce the growth rate.  The growth of these short-wavelength modes is primarily inhibited by resistive dissipation, because of the large $k$ required to reduce the stable stratification by such a large factor; typically, $k^2\eta\sim \Omega$. Therefore, strongly negative radial shear is required to drive these modes. Note that for very precise orientations of $k$ such that $|\kdotva|\ll 0.001 \Omega $ essentially hydrodynamic modes with similar $k$ may exist; however, these modes do not grow at a significantly faster rate than PSS modes, and they represent a very small portion of the phase space. As we show in Section \ref{sec:fielddependence}, however, these hydrodynamic modes become more important for smaller initial field magnitudes. Both the PSS mode and the CMRI mode are small-scale shear modes; these may be properly referred to as MRI modes.

For the radial magnetic field geometry, this region of phase space only has growing modes at colatitudes $9 \degrees < \theta < 45 \degrees$,  which correspond to shears of $d\ln \Omega/d \ln \varpi<-0.25$.  We find that for $r=0.692~\rr$, $\theta=30\degrees$, the most rapidly growing mode in this region of phase space has a growth rate $\Gamma=0.63\Omega$, and is located at the coordinates $k_\varpi/k_Z\approx-1.57$, $|\kdotva|=0.32\Omega$.  At latitudes where field-parallel modes in the second region are present, their growth rate is larger than that of nearly vertical modes in the first region by a factor of $\sim 10$; at $r=0.692~\rr$, $\theta=30\degrees$, the ratio of the growth rates is approximately $8.6$.

\subsubsection{Convectively Dominated Region C} \label{sec:phasec}

\begin{figure}[h]
\begin{center}
\includegraphics[width = .45\textwidth]{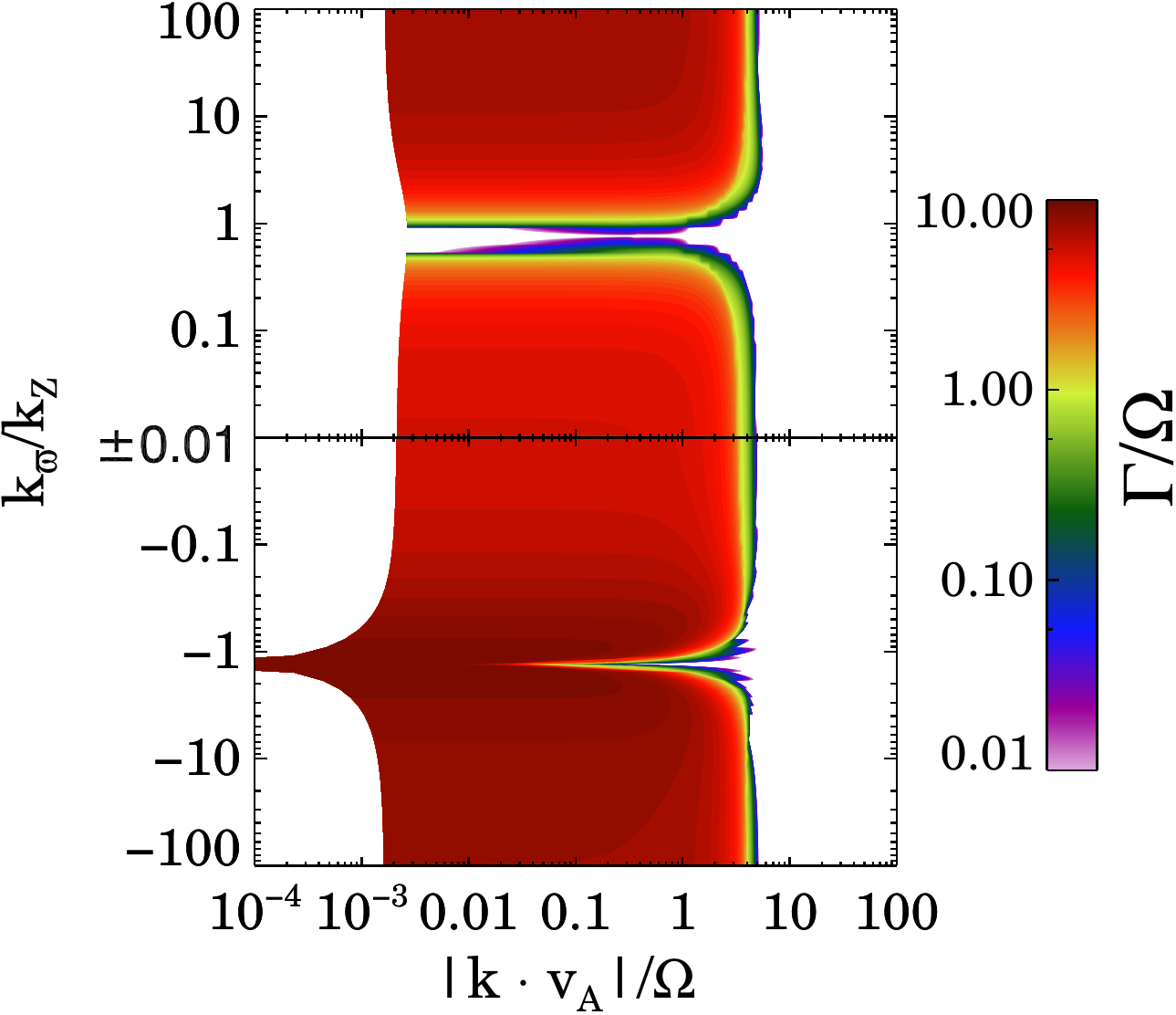}
\end{center}
\caption{{\small The dependence of the growth rate $\Gamma$ on the phase space parameters $k_\varpi/k_Z$ and $\left|\kdotva\right|$ at the location $r=0.965 ~\rr$, $\theta=37.5\degrees$ in Region C (Figure 2); the growth rate and $\left|\kdotva\right|$ are given in units of the local rotation rate $\Omega$. The most rapid growth is found at $k_\varpi/k_Z=-1.31$ and $|\kdotva|=7.8\times10^{-6}\Omega$, which corresponds to a hydrodynamic convective mode.}\label{fig:phasec}}
\end{figure}

In region C, growing modes are dominated by convection, while shear plays little role in the growth of instability, especially near the top of the convection zone. In Figure \ref{fig:phasec}, we show the dependence of the growth rate $\Gamma$ of the dispersion relation (\ref{eq:naxidisp}) on the parameters  $k_\varpi/k_Z$ and $\left|\kdotva\right|$ at the location $r=0.965~\rr$, $\theta=37.5\degrees$ in the upper part of Region C. The most rapid growth at this location corresponds to $k_\varpi/k_Z=-1.31$ and $|\kdotva|=7.8\times10^{-6}\ \Omega$, with maximum growth rate $\Gamma=10.4 \Omega$, similar to the local magnitude of the buoyancy frequency $|N|=10.5\Omega$. The dependence of the growth rate on $\left|\kdotva\right|$ is very weak over most of the phase space, with a slight decrease in growth rate at smaller scales; however, at critical values of $\kdotva$ corresponding to $k^2 \xi\approx N$, the growth rate abruptly drops from $\Gamma\approx N$ to $\Gamma=0$ with further increase in wavenumber. At $ k_\varpi/k_Z\sim -1.3$, the critical value of $\kdotva$ corresponding to this cutoff occurs at larger length scales because this orientation of the wavenumber corresponds to $|\kdotva|\ll k v_{\rm A}$; the critical value of $\kdotva$ still occurs at $k^2 \xi \sim N$. 

The variation of growth rate with $k_\varpi/k_Z$ in Region C is also relatively simple; there is a region of very low growth at $k_\varpi/k_Z\sim0.75$, while a region of rapid growth is found for $k_\varpi/k_Z\approx-1.3$, which is the approximate location of the most rapidly growing mode mentioned above. Because this variation is most evident for modes that are of very large length scale, we may identify the causes of this variation by using the dispersion relation Equation (\ref{eq:largescalemodes}) for such modes.  For the region under consideration at $r=0.965 ~\rr$,$\theta=37.5\degrees$, near the outermost layers of Region C, the appropriate regime is $\left| N\right| \gg \kappa,\Omega$. In this case, the growth rate $\Gamma_{\rm conv}=-i \omega$ of large scale convective modes is given by
\begin{equation}
\Gamma_{\rm conv}\approx \frac{|N|}{\sqrt{1+(k_\varpi/k_Z)^2}} \left|\frac{k_\varpi}{k_Z}\cos \theta-\sin \theta\right|, \label{eq:convgrowthc}
\end{equation}
The form of Equation (\ref{eq:convgrowthc}) indicates that, in general, we expect modes to grow faster for $k_\varpi/k_Z<0$, which is indeed what we calculate in the full analysis. 

To understand the regions of most rapid and slowest growth, we calculate the extrema of the growth rate given by Equation (\ref{eq:convgrowthc}). One extremum of the growth rate, $\Gamma_{\rm conv}$, occurs at $k_\varpi/k_Z=\tan \theta$, resulting in  $\Gamma_{\rm conv}=0$. For $r=0.965 ~\rr$, $\theta=37.5\degrees$, this extremum corresponds to $k_\varpi/k_Z=0.77$ and thus accounts for the region of low growth that we find at $k_\varpi/k_Z \approx 0.75$.  The second extremum of the growth rate occurs at $k_\varpi/k_Z=-\cot\ \theta$, which results in $\Gamma_{\rm conv}=|N|$. For the location discussed in this section, $\cot \theta=-1.31$; therefore, this analysis successfully predicts the orientation of the poloidal wavevector for the most rapidly growing mode. Note that the growth rate, $\Gamma$, found in our full analysis is typically smaller than $N$ because for $k_\varpi/k_Z=-\cot \theta$, ${\widetilde \kappa}>0$.

To better understand the physical reasons for the extrema given by Equation (\ref{eq:convgrowthc}), we use the constraint  $\delta v_\varpi/\delta v_z=-(k_\varpi/k_Z)^{-1}$ from Equation (\ref{eq6}) to calculate the orientation of the perturbed velocity flows for the two extrema.  Because convective instability has a maximum growth rate $\sim N$ and drives flows in the $r$ direction, we expect that the growth rate of convective modes will be $\Gamma\sim N \cos  \Phi_{r\delta v }$.
At $k_\varpi/k_Z=\tan \theta$ corresponding to vanishing growth rate, the perturbed flows have the ratio $\delta v_\varpi/\delta v_Z =- {\rm cot} \theta$, which corresponds to $N \cos  \Phi_{r\delta v }=0$. For $k_\varpi/k_Z=-\cot \theta$ corresponding to maximum growth rate, we find that $\delta v_\varpi/\delta v_z =\tan \theta$, which corresponds to a growth rate of $N \cos  \Phi_{r\delta v }=N$. These results for the two extrema are identical to those calculated using Equation (\ref{eq:convgrowthc}), and similar to the growth rates calculated from the full dispersion relation.

In this analysis, we have focused on the outermost part of Region C, where convection is completely dominant; however, for $r<0.9 ~\rr$ there are parts of phase space for which $k^2\xi >N$ and $|\kdotva| <\Omega$, in which convective effects are negligible and only shear can drive modes. These small--scale modes may have important effects on the nonlinear evolution of instability, even though they grow more slowly than the convective large--scale modes discussed here. 
 
\subsubsection{Region TL} \label{sec:phasetl}
 \begin{figure*}[t]
\begin{center}
\includegraphics[width = .95\textwidth]{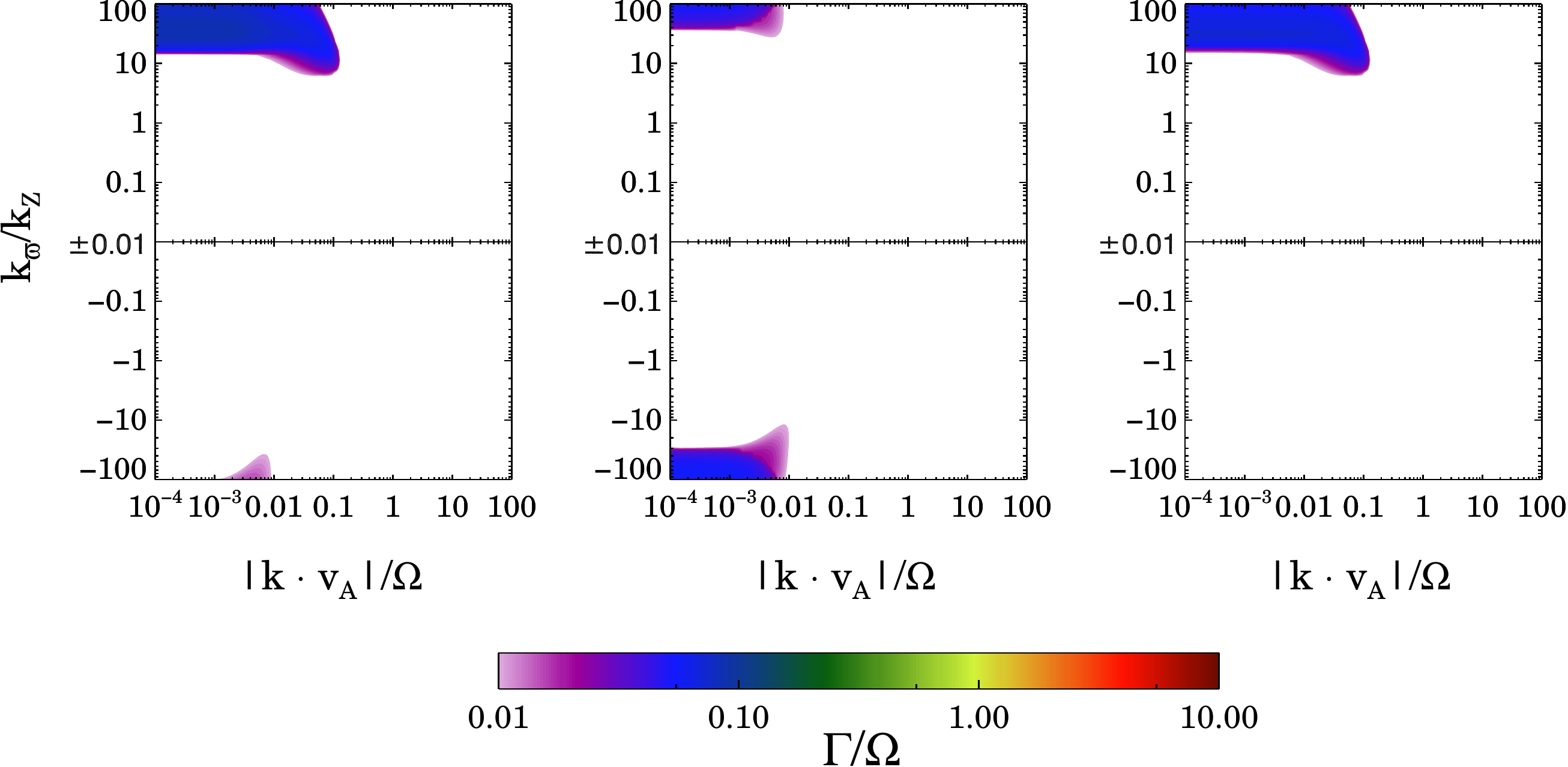}
\end{center}

\caption{{\small The dependence of the growth rate $\Gamma$ on the phase space parameters $k_\varpi/k_Z$ and $\left|\kdotva\right|$ at the location $r=0.721 ~\rr$, $\theta=75\degrees$ in Region TL including (left panel) the effects of both convection and shear, (middle panel) the effects of convection only, setting gradients of $\Omega$ to 0, (right panel) the effects of shear only, setting $N=0$. In all cases, the growth rate is given in units of the local rotation rate $\Omega$. The most rapid growth rate of unstable modes, which is significantly smaller than that in the other regions of the Sun, is found at $k_\varpi/k_Z=33.8$ and $|\kdotva|=5.27\times10^{-5}\ \Omega$.}\label{fig:phasetl}}
\end{figure*}In Region TL, the radial shear is positive, which typically inhibits the growth of modes. Nevertheless, modes can driven by either shear or convection in this region, as shown in Section \ref{sec:mrigrowth}.  Figure \ref{fig:phasetl} shows the dependence of the growth rate on the phase space parameters $k_\varpi/k_Z$ and $\left|\kdotva\right|$ at the location $r=0.721 ~\rr$, $\theta=75\degrees$ in Region TL, which is a location where the growth rates of shear modes and convective modes are approximately equal. Because $|\kdotva|<0.01 \Omega$ for all growing modes, the magnetic field has no significant effect in this region, and all modes are large--scale.   We find unstable modes only for $|k_\varpi/k_Z|>10$, which means that the unstable mode wavenumbers $\mathbf{k}$ are oriented nearly in the $\pm \varpi$ direction. Equation (\ref{eq6}) then shows that the corresponding unstable displacements are oriented nearly in the $\pm Z$ direction. One might expect that this occurs because positive radial shear inhibits motions in the $\varpi$ direction, but the middle panel of Figure \ref{fig:phasetl} shows that this inhibition exists even in the absence of shear.  For a large--scale mode with a displacement oriented in the $\pm\varpi$ direction, $k_\varpi/k_Z=0$, the dispersion relation (\ref{eq:largescalemodes}) becomes 

\begin{equation}
\omega^2-N^2 \sin^2 \theta -\kappa^2=0.
\end{equation}
Thus, in the absence of convection, the system will oscillate stably at the epicyclic frequency $\kappa$; we refer to the damping effects of these oscillations on growing modes as epicyclic stabilization. 

The maximum growth rate we find at this location in region TL is $\Gamma=0.096 \Omega$, which corresponds to the parameters $k_\varpi/k_Z=33.8$ and $|\kdotva|=5.27\times10^{-5}\ \Omega$; the growth rate is significantly smaller than the local values of $|N|$ and $\Omega$ in Region TL. Comparison of the left and middle panels of Figure \ref{fig:phasetl} reveals that the presence of positive radial shear significantly reduces the growth rate of convective modes with negative $k_\varpi/k_Z$, while producing growing modes with positive $k_\varpi/k_Z$ that, for this location, grow slightly faster than convective modes. It is important to note that because the radius of the tachocline decreases at large $\theta$, it is possible that no part of the tachocline is in the convectively unstable region with $r>0.713 ~\rr$ at latitudes corresponding to Region TL \citep{BA01}. If so, the effects of convection will be dominant throughout Region TL and shear will be significantly smaller at the bottom of Region TL, but the typical most rapidly growing modes will have similar growth rates to those calculated in this section.

Because all growing modes are large--scale, we can again make use of Equation (\ref{eq:largescalemodes}) to detail their properties. The growth rate of large--scale convective modes in Region TL may be calculated by setting gradients of $\Omega$ to 0 in Equation (\ref{eq:largescalemodes}). The resulting growth rate $\Gamma_{\rm conv}=-i \omega$ is then given by

\begin{equation}
\Gamma_{\rm conv}\approx \sqrt{-\frac{ 4\Omega^2 +([k_\varpi/k_Z] \cos \theta-\sin \theta)^2 N^2}{{1+(k_\varpi/k_Z)^2}}}. \label{eq:convgrowthtl}
\end{equation}

\noindent Again, the $4\Omega^2$ term represents the influence of epicyclic stabilization, since in the absence of shear $\kappa^2=4\Omega^2$. Growing convective modes exist only for 
\begin{equation}
\left|\frac{k_\varpi}{k_Z}-\tan \theta\right|> \frac{2 \Omega}{|N| \cos \theta}. \label{eq:convcondtl}
\end{equation} 

The growth rate $\Gamma_{\rm shear}\equiv -i \omega$ of modes driven by shear in the tachocline can then be found by setting $N=0$ in Equation (\ref{eq:largescalemodes}). In the tachocline, the shear is oriented approximately in the spherical $r$ direction, so

\begin{equation}
{\widetilde\kappa}^2\approx 2 \Omega^2\left(2+ q(\sin^2 \theta -\frac{k_\varpi}{k_Z}\cos \theta \sin \theta)\right),
\end{equation}

where $q\equiv d \ln \Omega/d \ln r$. Therefore, the growth rate of shear modes is given by

\begin{equation}
\Gamma_{\rm shear}\approx\sqrt{-\frac{4 \Omega^2+ 2q\Omega^2(\sin^2 \theta -k_\varpi/k_Z\cos \theta \sin \theta)}{1+(k_\varpi/k_Z)^2}}.\label{eq:shearlimit}
\end{equation}

The $4 \Omega^2$ term in the numerator of Equation (\ref{eq:shearlimit}) again represents the influence of epicyclic stabilization in the absence of shear. In Region TL, $q>0$, so growing shear modes exist only if 

\begin{equation}
\frac{k_\varpi}{k_Z}>\tan \theta +\frac{2}{q \sin \theta \cos \theta}. \label{eq:shearcondtl}
\end{equation}

 Neither convective modes nor shear modes can grow for $|k_\varpi/k_Z|< \tan \theta$; this quantitatively shows why growing modes do not exist in Region TL except at large values of $|k_\varpi/k_Z|$, where $\tan\theta>1$. At the location $r=0.721 ~\rr$, $\theta=75\degrees$, the dimensionless shear is given by $q=1.025$ and the local buoyancy frequency has magnitude $|N|=0.244 \Omega$. Equation (\ref{eq:shearcondtl}) therefore predicts that growing shear modes must have $k_\varpi/k_Z>11.5$, while Equation (\ref{eq:convcondtl}) predicts that growing convective modes must have $k_\varpi/k_Z>35.4$ or $k_\varpi/k_Z<-27.9$. The right and middle panels of Figure \ref{fig:phasetl} reveal that these conditions are indeed obeyed in the full analysis of shear and convective modes. 

\begin{figure*}[t]
\includegraphics[width = .95\textwidth]{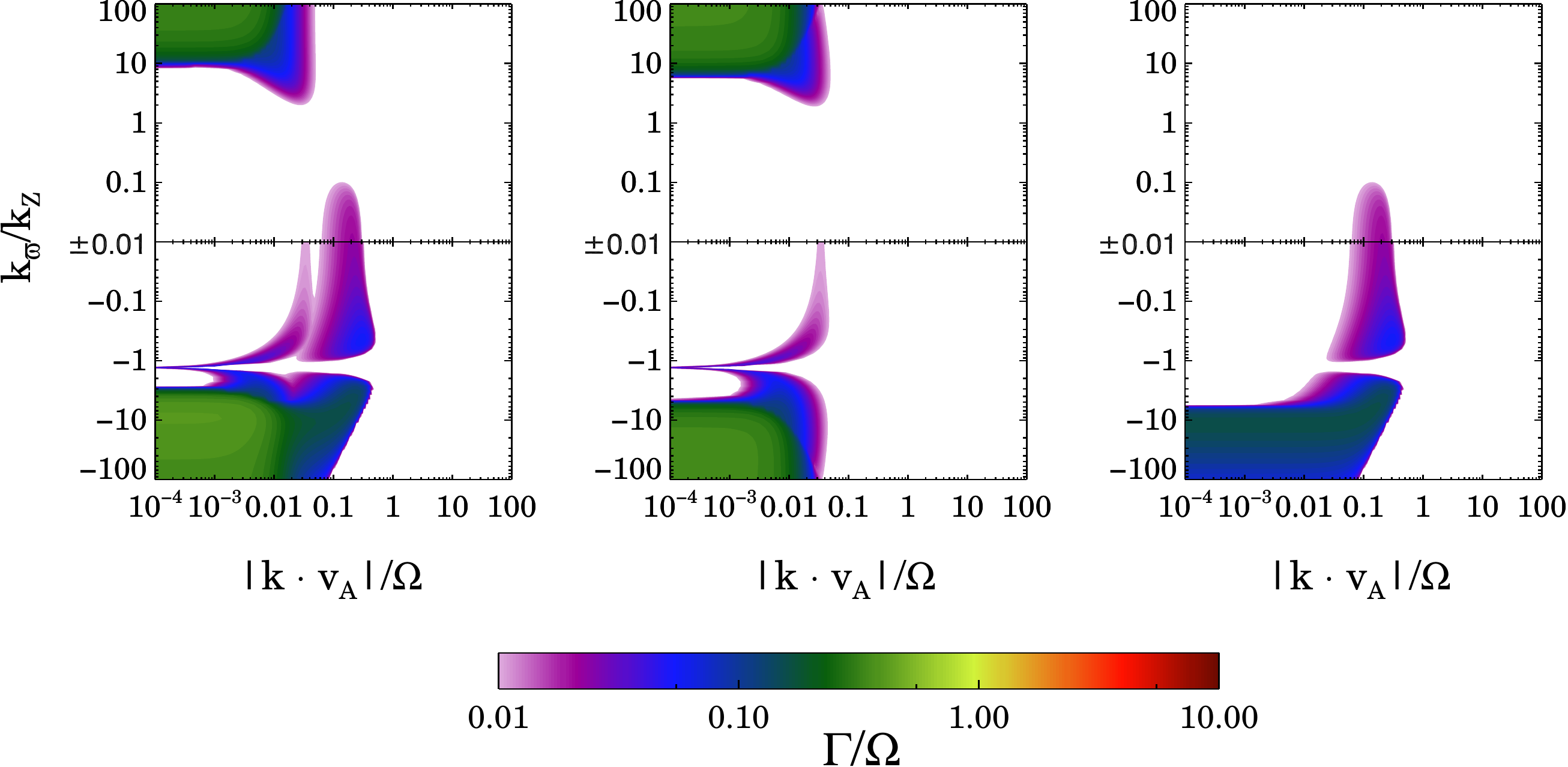}
\caption{{\small The dependence of the growth rate, $\Gamma$, on the phase space parameters $k_\varpi/k_Z$ and $\left|\kdotva\right|$ at the location $r=0.749~\rr$, $\theta=37.5\degrees$ in Region TU (Figure 2) including (left panel) the effects of both convection and shear, (middle panel) the effects of convection only, with gradients of $\Omega$ set to 0, (right panel) the effects of shear only, setting $N=0$. In all cases, the growth rate is given in units of the local rotation rate $\Omega$. The figure's middle and right panels reveal the presence of two shear modes and two convective modes, which are discussed in the text. The most rapid growth rate is $\Gamma=0.43 \Omega$ at the coordinates $k_\varpi/k_Z=-9.56$ and $|\kdotva|=4.0\times10^{-5}\ \Omega$; the growth is driven by both convection and shear.}\label{fig:phasetunotach}}
\end{figure*}
In order to understand why the processes that drive growing modes vary with latitude in Region TL, we now consider the variation of the growth rates of these modes with latitude. In the limit $|k_\varpi/k_Z|\gg 1$ applicable to modes in Region TL,  the convective growth rate given by Equation (\ref{eq:convgrowthtl}) becomes $\Gamma_{\rm  conv}\approx |N| \cos \theta$; therefore, the growth rate of convective modes decreases sharply as $\theta$ increases. In contrast, the growth rate of shear modes in Region TL does not vary sharply with $\theta$, although it does increase slightly with $\theta$ over most of region TL because both the radial shear $q$ and the rotational frequency $\Omega$ increase with $\theta$. Near the equator, however, the growth rate of shear modes decreases again, because less and less of the radial shear is oriented in the $Z$ direction, reducing the destabilization resulting from the condition given in Equation (\ref{eq:shearzmodes}). As a result, the overall growth rate falls below $0.05\Omega$ for $\theta>83 \degrees$.  Because of the suppression of all growing modes near the equator of the Sun, we expect that the production of large--scale features will be inhibited near the equator, especially at $\theta>83 \degrees$; we discuss the implications of this result for solar activity in Section \ref{sec:activeregions}.

\subsubsection{Region TU}

In Region TU the variety of growing modes is greater than in other regions. There are two important cases for which the growing modes are significantly different. In the upper tachocline at $r=0.721 ~\rr$, the shear is is negative and very strong, $|q| \gg 1$, while the buoyancy frequency is small compared to the rotation frequency $N \ll \Omega$. Thus, we expect that modes will be driven by shear. At slightly larger radii, shear and convection both contribute significantly to the growth of modes, and both are of the same order as $\Omega$. We will first discuss the case where both convection and shear are important, and then discuss how the results change in the shear--dominated case.

 Figure \ref{fig:phasetunotach} shows the dependence of the growth rate $\Gamma$ of the dispersion relation Equation (\ref{eq:naxidisp}) on the parameters  $k_\varpi/k_Z$ and $|\kdotva|$ at the location $r=0.749 ~\rr$, $\theta=37.5\degrees$, which is above the tachocline. The left panel shows the growth rate including the effects of both shear and convection, while the middle and right panels show the growth rate including only convection (setting gradients of $\Omega$ to 0), and including only shear (setting $N=0$). The figure reveals that there are two types of modes driven by shear and two types of modes driven by convection in Region TU. The two types of shear mode shown in the right panel are a large--scale hydrodynamic mode with a wide range of values of $|\kdotva|$ and $k_\varpi/k_Z <-10$ and a small-scale MHD mode with $0.025 \Omega<|\kdotva|<0.5 \Omega$ and  $-1.0<k_\varpi/k_Z <0.1$. The two types of convective mode shown in the middle panel are a mode with large $|k_\varpi/k_Z|$ and $k^2\xi< N$ and a highly overstable mode with oscillatory frequency much greater than its growth rate that corresponds to $k^2 \xi \approx N$ and exists at all values of $k_\varpi/k_Z<0$.  The two large--scale modes are very similar to the convective and shear modes discussed in Section \ref{sec:phasetl}; the sole difference is that in Region TU, $q<0$, so the large--scale shear mode grows only for $k_\varpi/k_Z<0$ and inhibits growth for $k_\varpi/k_Z>0$.  The most rapid growth rate found in the full analysis is $\Gamma=0.432 \Omega$, which occurs at the coordinates $k_\varpi/k_Z=-9.58$ and $|\kdotva|=4.0\times10^{-5}\ \Omega$. The instability at this location is driven by both convection and shear; the growth rate including only convection is $0.339\Omega$, while the growth rate including only shear is $0.168 \Omega$.
 
The small--scale shear mode in Region TU is similar to the CMRI mode discussed in Section \ref{sec:phasets}, and it can be referred to as an MRI mode. In Region TU this mode exists for all $\theta< 60 \degrees$, which corresponds to the region where $d\ln \Omega/d \ln \varpi<0$. The peak growth rate for this MRI mode at the location $r=0.749~\rr$, $\theta=37.5\degrees$ is $\Gamma=0.061 \Omega$, approximately $36\%$ of the growth rate of the hydrodynamic shear mode, and occurs at the coordinates $k_\varpi/k_Z=-0.44$ and $|\kdotva|=0.30\ \Omega$. The overstable convective mode at $k^2 \xi=N$ has maximum growth rate $\Gamma =0.048 \Omega$, while its oscillatory frequency is $\approx 1.05\Omega$. 
  
\begin{figure}[h]
\begin{center}
\includegraphics[width = .45\textwidth]{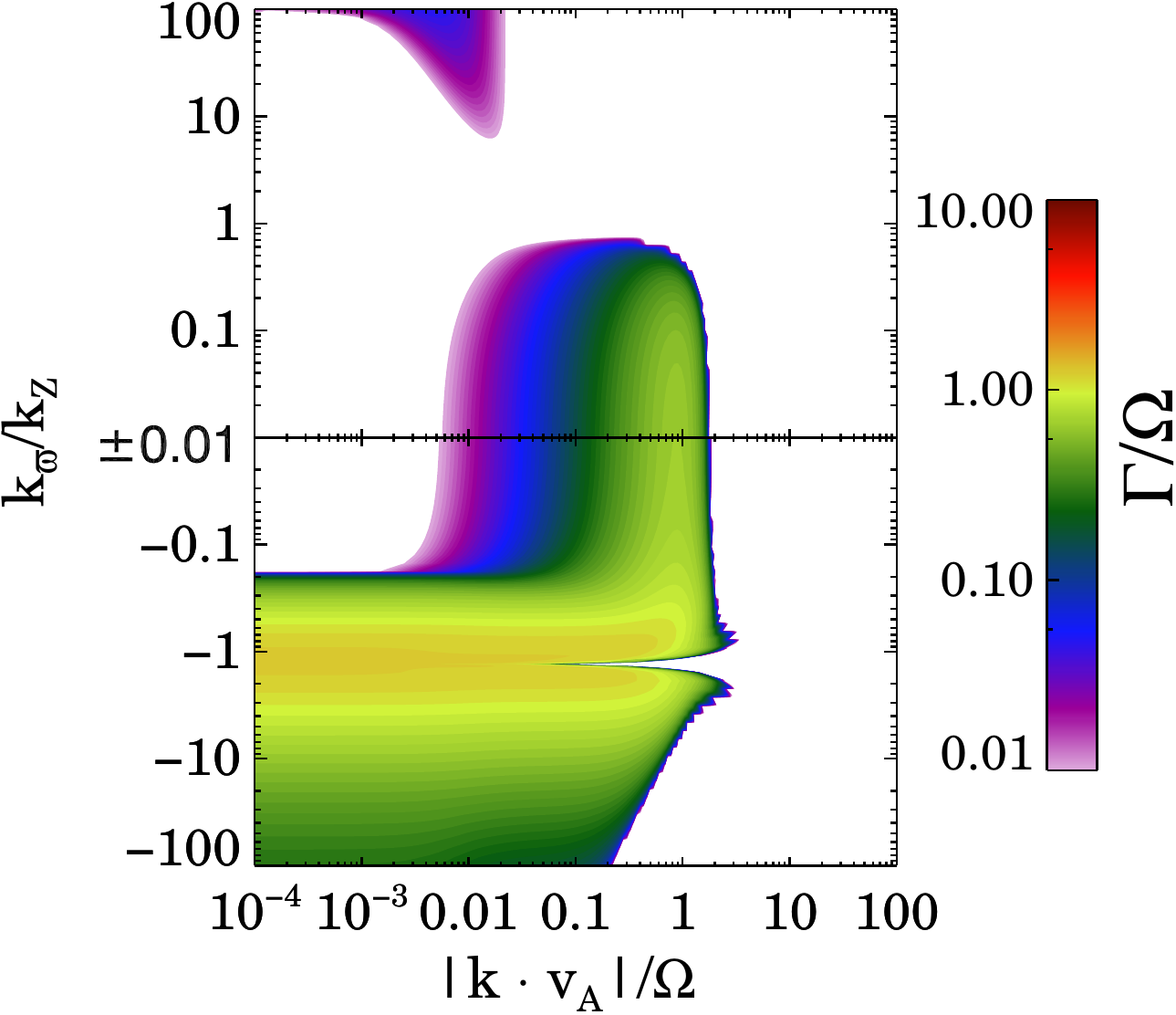}
\end{center}

\caption{{\small The dependence of the growth rate, $\Gamma$, on the phase space parameters $k_\varpi/k_Z$ and $\left|\kdotva\right|$ at the location $r=0.721 ~\rr$, $\theta=37.5\degrees$ in Region TU (Figure 2); the growth rate is given in units of the local rotation rate $\Omega$. The most rapid growth is found at $k_\varpi/k_Z=-1.202$ and $|\kdotva|=2.26 \times 10^{-6}\Omega$, which corresponds to a hydrodynamic shear mode.}\label{fig:phasetutach}}
\end{figure}

  We now discuss the growth rate of modes in the convectively unstable portion of the tachocline, where shear is dominant and thermal buoyancy is very weak.   Figure \ref{fig:phasetutach} shows the dependence of the growth rate $\Gamma$ of the dispersion relation Equation (\ref{eq:naxidisp}) on the parameters  $k_\varpi/k_Z$ and $\left|\kdotva\right|$ at the location $r=0.721 ~\rr$, $\theta=37.5\degrees$. In this figure, the two shear modes have merged, and their growth rate is significantly larger due to the strong shear in the tachocline.  The convective modes have been swamped by the shear modes, except at $k_\varpi/k_Z>10$, where a convective mode is present but strongly suppressed compared to the growth rate that it would have in the absence of shear. 

The most rapidly growing mode at this location has a growth rate of $1.38 \Omega$, and occurs for parameters $k_\varpi/k_Z=-1.202$, $|\kdotva|=2.26 \times 10^{-6} \Omega$. These parameters correspond to a large--scale hydrodynamic shear mode; because the effect of convection is negligible compared to that of shear, its growth rate is 
\begin{equation}
\Gamma_{\rm shear}\approx \Omega\sqrt{-\frac{4+ 2q(\sin^2 \theta -[k_\varpi/k_Z]\cos \theta \sin \theta)}{1+(k_\varpi/k_Z)^2}}, 
\end{equation}

For the location under consideration at $r=0.721 ~\rr$, $\theta=37.5 \degrees$, $q=-4.446$. The only extremum that corresponds to a growing mode occurs at $k_\varpi/k_Z=-1.18$, which gives $\Gamma_{\rm shear}=1.35\Omega$. These values are close to those of the most rapidly growing mode with $k_\varpi/k_Z=-1.20$ and growth rate $\Gamma=1.38$ calculated in the full analysis. It is important to note that in the convectively unstable tachocline the MRI mode at low $|k_\varpi/k_Z|$ grows nearly as fast as does the large--scale  hydrodynamic shear mode. Thus, MRI modes are likely to be important in both Region TU and Region TS.

\subsubsection{ Effect of Field Geometry on Axisymmetric Modes}\label{sec:fielddependence}

We find that neither hydrodynamic shear modes nor convectively driven modes are strongly affected by the field geometry, except in the case, unrealistic for the Sun, where the magnitude of the magnetic field is large enough that it becomes dynamically important. The apparent dependence on $\kdotva$ for such modes is merely a dependence on $k$. In contrast, the small--scale shear modes found in Regions TS and TU, which are true MRI modes, have growth rates that depend on the field geometry. For small--scale modes, $\waphi\ll \kdotva$, so the growth rates depend only on the poloidal magnetic field ratio $B_\varpi/B_Z$ and the magnitude of the poloidal field $B_{\rm pol}$.  We now calculate maximum growth rates for each type of MRI mode in Regions TS and TU.  We choose values of $B_\varpi/B_Z$ corresponding to a field oriented in the $r$, $\theta$, and $Z$ directions\footnote{We do not carry out an detailed analysis using fields oriented in the $\varpi$ direction because the CMRI and PSS modes overlap with each other for this field orientation; this significantly complicates the analysis.} and field magnitudes ranging from $10^{-4}$ G to $10^4$ G; in the large--field limit, we include magnetic field values similar to those in the present--day Sun. In order to isolate the CMRI mode and the MRI mode in Region TU, which can overlap with other modes, we calculate their maximum growth rates only for the region of phase space with $|k_\varpi/k_Z|<0.1$.

The first type of small--scale shear mode is the CMRI mode in Region TS, which typically corresponds to small $|k_\varpi/k_Z|$. At the location $r=0.692 ~\rr$, $\theta=30\degrees$ (Figure 2), we find that for a poloidal field magnitude of $0.2$ G, the maximum growth rates for this type of mode are $0.073 \Omega$, $0.142 \Omega$, and $0.063 \Omega$ for $B$ oriented in the $r$, $\theta$, and $Z$ directions, respectively. CMRI modes with growth rate larger than $0.01 \Omega$ exist for the CMRI mode only in the narrow range of poloidal field magnitudes $0.02$ G$<B_{\rm pol}<0.6 $ G. For larger magnetic fields, all modes with $k$ large enough to reduce the effects of stratification have $|\kdotva|\gg \Omega$; as a result, the magnetic tension prevents the growth of instability. In contrast, for smaller fields the condition $\kdotva \sim \Omega$ for small-scale shear modes implies that $k$ is very large; the growth of these modes is then strongly inhibited by resistive and viscous dissipation, because the characteristic dissipative frequencies are proportional to $k^2$.

\begin{figure}[h]
\begin{center}
\includegraphics[width = .45\textwidth]{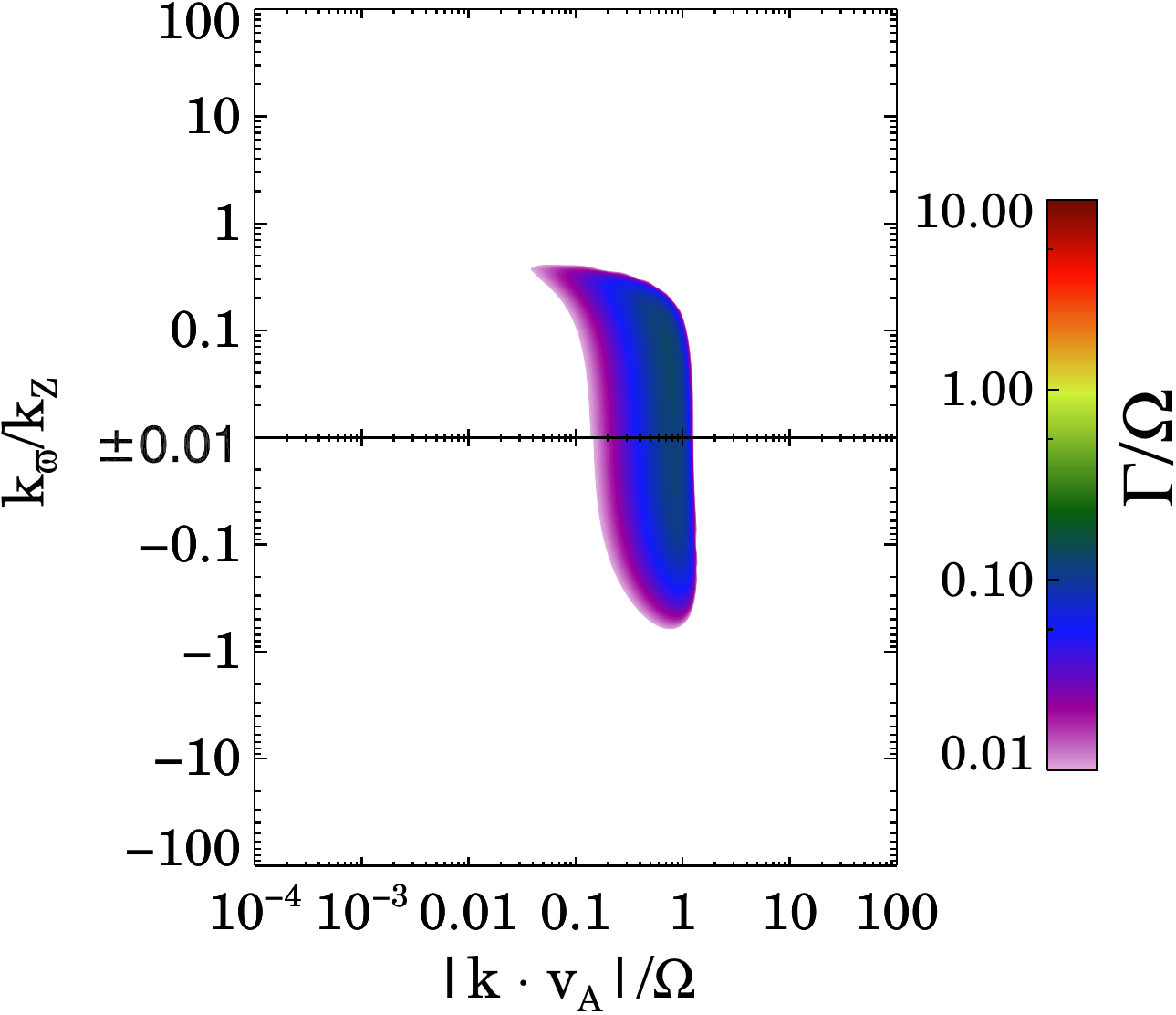}
\end{center}

\caption{{\small The dependence of the growth rate, $\Gamma$, on the phase space parameters $k_\varpi/k_Z$ and $\left|\kdotva\right|$ at the location $r=0.692 ~\rr$, $\theta=30\degrees$ in Region TS (Figure 2) for a magnetic field oriented in the $\theta$ direction. Only the CMRI mode at small $|k_\varpi/k_Z|$ is present, while there is no mode corresponding to $|\kdotva|\ll k v_A$ (Compare to Figure \ref{fig:mrigrowth}). The most rapid growth is found at $k_\varpi/k_Z=0.068$ and $|\kdotva|=0.753\Omega$, which corresponds to a CMRI mode.}\label{fig:phasetstheta}}
\end{figure}

The second type of mode found in Region TS is the PSS mode with $|\Phi_{kB}|\approx 90 \degrees$ and $|\kdotva|\ll k v_A$; for PSS modes, the stabilizing effects of magnetic tension are substantially reduced. At the location $r=0.692 ~\rr$, $\theta=30\degrees$, this mode grows quickly only for relatively large poloidal field strengths $0.08$ G~$<B_{\rm pol}$, and the growth rate depends strongly on the orientation of the field.  For a poloidal field of $0.2$ G the growth rate of this mode is $0.63 \Omega$ for a magnetic field oriented in the $r$ direction, while the growth rate for modes oriented in the $\theta$ and $Z$ directions is smaller than $0.1 \Omega$ and represents an extension of the CMRI mode. Figure \ref{fig:phasetstheta} shows the dependence of the growth rate on the phase space parameters $k_\varpi/k_Z$ and $\left|\kdotva\right|$ at the location $r=0.692 ~\rr$, $\theta=30\degrees$ for a magnetic field oriented in the $\theta$ direction, revealing the presence of the CMRI mode and the absence of the PSS mode (contrast to the structure in Figure \ref{fig:phasets}). Comparing the growth rates for various field orientations $B_\varpi/B_Z$ indicates that the most rapid growth occurs for $\Phi_{Br}\approx 0 \degrees$, but with field orientation shifted towards $B_\varpi/B_Z = 1$. At  $r=0.692 ~\rr$, $\theta=30\degrees$, the PSS mode grows most quickly for $B_\varpi/B_Z = 0.84$, which corresponds to $\Phi_{Br}\approx 10 \degrees$; growth rates are significantly larger than those for the CMRI mode for orientations within $ \pm 45 \degrees$ of this orientation that gives the most rapid growth. The range of orientations for most rapid growth is typically similar at other latitudes. For larger $B_{\rm pol}\gg1$ G, the PSS mode grows at similar rates to those found for $B_{\rm pol}=0.2$, but the range of wavenumber orientations for which growth occurs decreases significantly. The pure PSS mode does not grow appreciably for small magnetic fields, but a very small--scale hydrodynamic mode becomes important for $B_{\rm pol}<0.05$ G; this mode has $k_\varpi/k_Z$ that is similar to that of the PSS mode, and the growth rate is quite large, $\sim 0.67 \Omega$.  For this hydrodynamic PSS mode, the field orientation is unimportant, because the small field means the magnetic tension does not reduce the growth rate of modes.  

The strong dependence of growth rates on field orientation for PSS modes may be explained as follows: for fields oriented in the $\theta$ and $Z$ directions, the constraint on PSS modes $|\kdotva|\ll k v_A$, which corresponds to $|\Phi_{kB}|\approx 90 \degrees$, implies that $|\Phi_{kr}|\lta 45 \degrees$. Because all MRI modes are driven by negative radial shear, these PSS modes will experience significant rapid growth only for  $|\Phi_{kr}|\approx 90 \degrees$, which corresponds to radial displacement being the cause of instability; however, parameters for which shear drives growing modes do not coincide with parameters for which $|\kdotva|\ll k v_A$. As shown in Section \ref{sec:phasets}, this type of mode typically grows faster than the CMRI mode, so approximately radially oriented fields lead to the most rapid MRI growth rates.
  
In Region TU, we find that no significant growth of modes with $|\kdotva|\sim \Omega$ occurs for initial magnetic fields of magnitude smaller than $\sim 0.002 $ G, but growth can occur even for large magnetic fields $B\sim 10^4 $ G similar to those in the present--day Sun. The growth of modes with $|\kdotva|\sim \Omega$ is inhibited by resistive and viscous dissipation for small fields, but for large fields the absence of stable stratification means that modes for which $|\kdotva|\sim \Omega$ can grow even though they correspond to relatively large mode wavelengths. For these large wavelengths, the most rapidly growing modes become adiabatic because their wavelengths are large enough that thermal diffusion is negligible. 

  The growth rate of the MRI mode in Region TU at the location $r=0.721 ~\rr$, $\theta=37.5\degrees$ for $B_{\rm pol}=0.2$ G is approximately $0.85 \Omega$, and it varies by less than $5\%$ with orientation. The growth rate does not change significantly for larger magnitudes of $B$, and the dependence on field orientation remains very small. In contrast, the growth rate decreases quickly as $B$ is decreased, dropping to $\sim0.01 \Omega$ for $B_{\rm pol}=0.004 $ G; the decrease in growth rate is similar for the MRI mode.  The dependence of growth rates on orientation is also significantly increased for small $B$; the growth rate for a field oriented in the $Z$ direction is typically approximately twice that for a field oriented in the $r$ and $\theta$ directions for $B_{\rm pol}<0.02$ G.    

\subsection{Nonaxisymmetric Effects}\label{sec:naxieffects}

We now consider the effects of nonaxisymmetry on the growth rate of modes. Because all nonaxisymmetric terms in the dispersion relation are proportional to $\kdotva$ or $\waphi$, they are typically negligible for large--scale hydrodynamic shear modes and for all types of convective modes if $m$ is small. For very large $m$, the growth rate of these large--scale modes is slightly reduced by the toroidal magnetic tension, but this effect is very small. For the small--scale MRI modes in Regions TS and TU, the effects of nonaxisymmetry are considerably stronger and more complex. In general, we find that the most rapidly growing MRI modes are always nonaxisymmetric in these regions, although the differences in growth rate are very small for the maximum nonaxisymmetric wavenumber $|m|=15$ used in Section \ref{sec:mrigrowth}.

 For MRI modes in regions of the Sun close to the tachocline with $\varpi \sim 0.7~\rr$ $\Omega\sim 2 \times 10^{-6} \ {\rm rad/s}$, $\rho\sim0.2 {\rm g \ {cm}^{-3}}$, and the small seed fields $B \sim 1 G $ and moderate ratios of toroidal to poloidal field $|R_{\rm TP}|\sim 5$ that we have used in our analysis, $\waphi \sim 10^{-5} \Omega$. Therefore, the explicitly nonaxisymmetric terms in Equation (\ref{eq:naxidisp}) are negligible, and nonaxisymmetric effects derive almost entirely from the contribution $m \waphi$ of the nonaxisymmetric field and wavenumber to $\kdotva$. Because $(\kdotva)_{\rm pol}\sim \Omega$ for MRI modes, we can parameterize the strength of nonaxisymmetric effects using the parameter $\mu\equiv m \waphi/\Omega$. In the solar tachocline, this corresponds to $\mu\sim  10^{-5}m$. Therefore, our calculation with $|m|<15$ corresponds to $\mu\ll 1$, and nonaxisymmetric effects may be treated as small perturbations to the axisymmetric dispersion relation (\ref{eq:axidisp}). In this linear regime, the growth rate varies linearly with $\mu$, and the most rapidly growing modes will always be nonaxisymmetric with the highest possible value for $|m|$; this is exactly what we have found in our calculation of growth rates in Section \ref{sec:mrigrowth}. 

While it is relatively easy to predict that a nonaxisymmetric mode of maximal $|m|$ will be the most rapidly growing mode, it is extremely difficult to explain analytically which sign for $B_Z k_Z/(m \waphi)$ leads to faster growth because $\kdotva$ is present throughout the dispersion relation; we do not attempt to do so in this paper. Numerically, we find that for the CMRI mode, the sign of $B_Z k_Z/(m \waphi)$ that gives the smaller value of $|\kdotva|$ yields the most rapid growth rate. For PSS modes, the same is typically true, although the sensitivity of the dependence on $B_Z k_Z/(m \waphi)$ is typically very small.  In contrast, for the mode in Region TU, the sign of $B_Z k_Z/(m \waphi)$ that increases $|\kdotva|$ always leads to the largest growth rate.  For modes in Region TS, the value of $k^2$ for which modes grow most rapidly is determined by the stable stratification for the CMRI mode, and by resistive dissipation for the PSS mode. Therefore, the reduction in magnetic tension for a given value of $k$ may mean that the mode can grow slightly faster. For the small--scale mode in Region TU, the magnitude of $|N|$ is very small  and $k$ is moderate in magnitude, so magnetic effects are responsible for both driving the instability and stabilizing modes via magnetic tension. On balance, the driving of modes resulting from slightly increased $|\kdotva|$ for a given $k$ appears to be more important than the increase in magnetic tension.
\begin{figure}[t]
\begin{center}
\includegraphics[width = .45\textwidth]{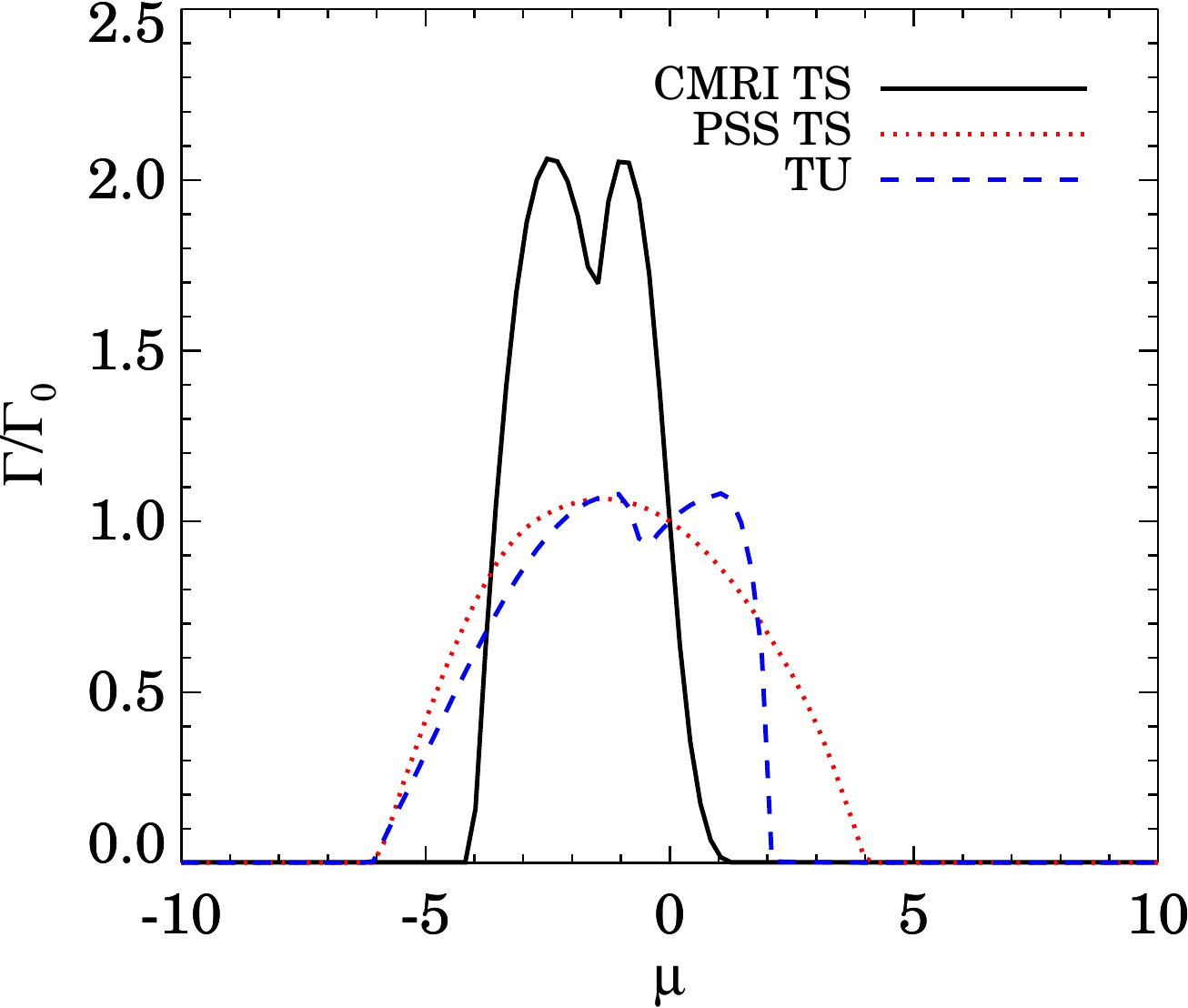}
\end{center}
\caption{{\small The dependence on $\mu$, $\mu$, the normalized nonaxisymmetric component of $\kdotva$, of the ratio of the growth rate, $\Gamma$, of the most rapidly growing nonaxisymmetric modes with dimensionless toroidal wavenumber, $\mu$, to the growth rate, $\Gamma_0$, of the most rapidly growing axisymmetric modes. The curves correspond to the specific ratio of toroidal to poloidal field $R_{\rm TP}=500$. The three types of modes shown are the CMRI and PSS modes in Region TS at $r=0.692 ~\rr$, $\theta=30 \degrees$ and the MRI mode in Region TU at $r=0.721 ~\rr$, $\theta=37.5 \degrees$ (Figure 2).}\label{fig:naxigrowth500}}
\end{figure}

\subsubsection{Nonaxisymmetric Mode Growth Rates}
To ascertain the effects of nonaxisymmetry for small--scale modes, we now compare the growth rate $\Gamma$ of nonaxisymmetric modes with larger wavenumbers $|m|>15$ to the growth rate $\Gamma_0$ of the corresponding axisymmetric mode. The local approximation requires that $m/\varpi \ll k_{\varpi}, k_Z$, which corresponds to the condition on $\mu$
\begin{equation}
\mu\ll\varpi R_{\rm TP} \ {\rm min}(1, |k_{\varpi}/k_Z|).\label{eq:localapproxnaxi}
\end{equation}

\noindent We therefore analyze nonaxisymmetric effects for toroidal--to--poloidal field ratio $R{\rm TP}=500$, which ensures that the local approximation is obeyed for modes with $|\mu|<10$.  This allows us to explore nonaxisymmetric effects at $\mu\gta1$ which are not present in the linear regime.

Figure \ref{fig:naxigrowth500} shows the dependence of the growth rate, $\Gamma$, of the most rapidly growing nonaxisymmetric modes on $\mu$, the normalized nonaxisymmetric component of $\kdotva$. The growth rate is given in terms of the growth rate, $\Gamma_0$, of the most rapidly growing mode with $\mu=m=0$. We use the same poloidal field geometry as in our axisymmetric analysis, with $B_{\rm pol}=0.2$ G and $B_{\varpi}/B_Z=\tan \theta$, which corresponds to a magnetic field oriented in the $r$ direction, and set the sign $B_Z k_Z/(\waphi)>0$. For $\mu \ll 1$, the size of nonaxisymmetric effects is small, and the variation with $\mu$ of the growth rate $\Gamma$ is approximately linear; this is in line with our predictions for the linear regime in the previous section. 
\begin{figure*}[t]
\begin{center}  
\includegraphics[width = .95\textwidth]{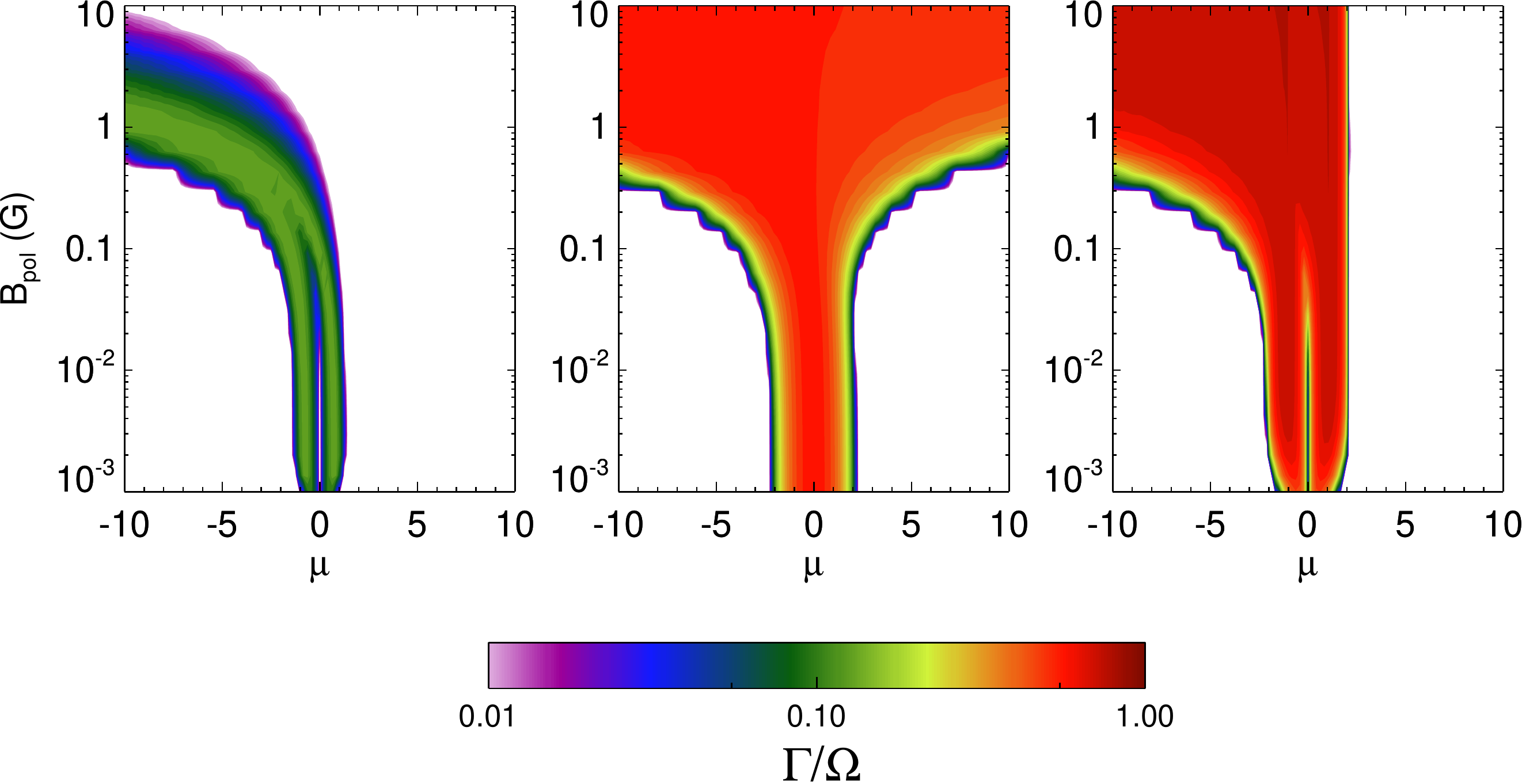}
\end{center}

\caption{{\small The dependence of the growth rate, $\Gamma$, of nonaxisymmetric MRI modes on the poloidal magnetic field $B_{\rm pol}$ and the dimensionless toroidal wavenumber $\mu$ for the CMRI  (left panel) and PSS mode (middle panel) in Region TS at $r=0.692 ~\rr$, $\theta=30 \degrees$ and the small--scale mode in Region TU (right panel) at $r=0.721 ~\rr$, $\theta=37.5 \degrees$. The increase in the range of $\mu$ with $B_{\rm pol}$ reflects the fact that because $R_{\rm TP}$ is held constant, $\mu \propto B_{\rm pol}$. }}\label{fig:naxibmag}
\end{figure*}

For $|\mu|\gta1$, the growth rate of nonaxisymmetric modes peaks and then goes down as the tension produced by the toroidal field begins to inhibit growth. The maximum value of $\Gamma$ corresponds to $\mu<0$ for both modes with $r=0.692~\rr$, $\theta=30 \degrees$ in Region TS, and $\mu>0$ for the mode in Region TU at $r=0.721 ~\rr$, $\theta=37.5 \degrees$. The sign of $\mu$ at each of these maxima is the same as the sign of $d\Gamma/d\mu$ found for each of these modes in Section \ref{sec:naxieffects} at $\mu \ll 1$ in the linear regime. The stabilizing effects of magnetic tension appear to be larger for a given $|\mu|$ if $\mu>0$ for all three modes.  This is because we have used the sign convention $(\kdotva)_{\rm pol}>0$, so that large positive $\mu$ corresponds to a larger value for $|\kdotva|$ and thus a larger magnetic tension than does large negative $\mu$. 

As shown in Figure \ref{fig:naxigrowth500}, the maximum growth rate of nonaxisymmetric modes is $2.07 \Gamma_0$ for the CMRI mode, $1.09 \Gamma_0$ for the mode in Region TU, and $1.07 \Gamma_0$ for the PSS mode.  For the CMRI mode and the mode in Region TU, two peaks are present in the growth rate, corresponding to values of $\mu$ that produce values of $\kdotva$ of the same magnitude but opposite sign. The locations of these peaks are offset from $\mu=0$ in the direction of increasing growth rate in the linear regime. For these modes, nonaxisymmetric effects thus correspond to an adjustment in $\kdotva$ that breaks its degeneracy with the magnitude of $k$, and the peaks correspond to the values of $|\kdotva|$ that are optimized for fast growth, rather than being determined by dissipative constraints on $k$. 

The sensitivity of the growth rate to $\mu$ is significantly weaker for the PSS mode than for the other two modes, and only one peak in growth rate is present.  This is a result of the strong dependence of the growth of PSS modes on $k_\varpi/k_Z$, which is required to ensure that the large magnitude of $k$ does not stabilize the mode via magnetic tension. When a nonaxisymmetric wavenumber resulting in $\mu\neq 0$ is introduced, its contribution to $\kdotva$ can be canceled by shifting the value of $k_\varpi/k_Z$ slightly; for $r=0.692~\rr$, $\theta=30 \degrees$ in Region TS, the value of $k_\varpi/k_Z$ is shifted from the axisymmetric value of $-1.57$ to $-1.20$ for the most rapidly growing modes at large $|\mu|$. Thus, $\mu$ is degenerate with the parameter $k_\varpi/k_Z$ for the PSS modes, and the degeneracy between $k$ and $\kdotva$ is not broken by nonaxisymmetric effects. This reduces the importance of nonaxisymmetric effects for PSS modes.  

\subsubsection{Effects of Field Geometry on Nonaxisymmetric Modes} \label{sec:naxifieldgeometry}
The appropriate field geometry in the nonaxisymmetric case is determined by three parameters: the direction of the poloidal field $B_\varpi/B_Z$, the toroidal--to--poloidal field ratio $R_{\rm TP}$, and the magnitude of the poloidal field $B_{\rm pol}$. We find that the effects of $B_\varpi/B_Z$ on the nonaxisymmetric MRI modes are similar to those on their axisymmetric counterparts; the PSS mode exists only for mode directions that are close to radial, while growth rates for the CMRI mode and the small--scale shear mode in Region TU vary by factor of 2 at most with direction. The initial toroidal--to--poloidal field ratio $R_{\rm TP}$ has been subsumed into the dimensionless wavenumber $\mu$; once this has been done, the only effect $R_{\rm TP}$ then has is in determining the range of $\mu$ for which our linear analysis is valid. The poloidal field magnitude $B_{\rm pol}$ is therefore the only field geometry parameter that has significant implications for the strength of nonaxisymmetric effects in our linear analysis. 

Figure \ref{fig:naxibmag} shows the dependence of the growth rate of the three MRI modes on the poloidal magnetic field $B_{\rm pol}$ and the dimensionless toroidal wavenumber $\mu$ for toroidal--to--poloidal field ratio $R_{\rm TP}=500$. The typical dependence on $\mu$ for all modes is similar to that described in the previous section, including a double peak in growth rate for the CMRI mode and the mode in Region TU, and a single peak for the PSS mode for most values of $B_{\rm pol}$. 

For the CMRI mode, which is shown in the left panel of Figure \ref{fig:naxibmag}, the nonaxisymmetric growth rate is larger than the axisymmetric growth rate by a very large factor of $>100$ for small initial poloidal field $B_{\rm pol}\sim 0.001$. As $B_{\rm pol}$ is increased, the ratio of the growth rates of the most rapidly growing nonaxisymmetric modes to those of axisymmetric modes eventually becomes smaller, reaching $\Gamma/\Gamma_0\sim2$ for $B_{\rm pol}\sim 1$ G. As $B_{\rm pol}$ is increased further, the poloidal magnetic tension begins to stabilize the growth of modes, as described in  Section \ref{sec:fielddependence}. For $B_{\rm pol}>1$ G, axisymmetric modes are stabilized completely, but modes with $\mu<0$ are able to grow because the nonaxisymmetric contribution to $\kdotva$ reduces the impact of the poloidal tension. 

For the mode in Region TU, shown in the right panel of Figure \ref{fig:naxibmag}, the nonaxisymmetric growth rate is again larger than the axisymmetric growth rate by a very large factor of $>100$ for small initial poloidal field $B_{\rm pol}\sim 0.001$. As $B_{\rm pol}$ is increased, the ratio $\Gamma/\Gamma_0$ of the growth rates of the most rapidly growing nonaxisymmetric modes to those of axisymmetric modes decreases quickly, falling below $1.2$ for poloidal fields $B_{\rm pol}\sim 0.2$ G; however, nonaxisymmetric modes remain the most rapidly growing modes. For even larger $B_{\rm pol}$, fast nonaxisymmetric growth occurs for large negative $\mu$, but not for large positive $\mu$; this is because negative $\mu$ reduces the magnetic tension.

The typical evolution of the PSS mode, shown in the middle panel of Figure \ref{fig:naxibmag}, is very different from that of the other two modes. For all values of $B_{\rm pol}$, the nonaxisymmetric and axisymmetric modes have very similar growth rates. For $B_{\rm pol}<10^{-2}$ the calculated PSS mode growth corresponds to the hydrodynamic mode discussed in Section \ref{sec:fielddependence}, and axisymmetric modes grow faster than nonaxisymmetric modes. For larger $B_{\rm pol}$, the calculated growth corresponds to the PSS mode proper, and nonaxisymmetric modes grow slightly faster than axisymmetric modes. The weak dependence of growth rate on $\mu$ for all $B_{\rm pol}$ reflects the degeneracy between $\mu$ and $k_\varpi/k_Z$ for the PSS mode. 

\section{Discussion}\label{sec:discussion} 
We now compare our results to other authors and discuss the nonlinear effects leading to the saturation of the magnetic field.

\subsection{Comparison with Other Studies}
\label{sec:mricomparison}
Our axisymmetric study of modes in the Sun is most directly comparable to the research of \citet{PM07} and \citet{M11}.  \citet{PM07} studied the growth of modes in the stably stratified tachocline. They found that significant growth of instability occurred for $\theta< 53 \degrees$, and that initially radial fields lead to faster growth than do toroidal fields over most of the domain. The trends we find are similar; however, by studying the phase space structure of the dispersion relation we have identified the two small--scale shear modes  in Region TS and found that the reason radial fields lead to faster growth is that they allow the fast growth of PSS modes.  Quantitatively, the growth rates found by \citet{PM07}  are somewhat larger than ours in Region TS because we use a larger value for $N$ corresponding to the bottom of the tachocline, while they use a smaller value corresponding to its center.

\citet{M11} investigated the growth of axisymmetric modes throughout the Sun in the absence of thermal buoyancy; thus, the only unstable modes present were MRI modes. He found that the MRI is unstable only near the tachocline at high latitudes and very close to the surface. The stabilizing effect of 
density stratification was substantially reduced by doubly--diffusive 
effects in the convectively stable portion of the tachocline. He found that the layer 
near the surface is formally unstable to the MRI, but argued that the
dispersion relation that is the basis of the instability criteria should 
not be applied in a region of vigorous convection since the stationary 
background assumed to derive the dispersion relation does not exist there. 
We have confirmed using a linear analysis of the full dispersion relation that modes driven by shear are not present in the outer parts of the convection zone, which corresponds to Region C. This is because for moderate seed fields $B\sim 1$ G, modes with $\kdotva\sim \Omega$ that correspond to the shear modes also have $k^2\xi< N$; therefore, hydrodynamic convective modes with these parameters have a growth rate $\sim |N|$, and there are no parameters for which shear modes are significant. 

Our nonaxisymmetric results can be most directly compared to those found by \citet{MSS07} in the context of stably stratified proto--neutron stars. They found that the most rapid growth of nonaxisymmetric modes occurred for $\mu\sim 1$, and that the nonaxisymmetric modes grew much faster than axisymmetric modes unless the poloidal field was very large. We find that the relative growth rates of nonaxisymmetric and axisymmetric modes depend on the type of mode. For the CMRI mode, the nonaxisymmetric modes always grow significantly faster than axisymmetric modes, especially for very large and very small poloidal fields.  For the PSS mode, nonaxisymmetric modes never grow more than 5\% faster than axisymmetric modes; this mode is not detected by \citet{MSS07} because their dispersion relation assumes that $B_{\rm \varpi}=0$. Finally, for the mode in the convectively unstable Region TU, nonaxisymmetric modes grow much faster than axisymmetric modes only for very small poloidal fields. \citet{MSS07} do not investigate the CMRI case with very large fields, for which nonaxisymmetric CMRI modes are dominant because the axisymmetric modes are stabilized by magnetic tension; therefore, our results are consistent within their range of validity.

\subsection{Hydrodynamic Modes and the Emergence of Active Regions}
\label{sec:activeregions}
Our results for Region TL suggest that the growth of large--scale modes is significantly suppressed by epicyclic stabilization at $ \theta> 83 \degrees$. \citet{PM07} suggested that MRI modes in the convectively stable tachocline suppress the formation of large--scale features for $\theta\le53 \degrees$.  Our results in Region TU suggest that MRI modes will also be important in the convectively \textit{unstable} tachocline for $53 \degrees \le \theta \le 60 \degrees$. The combination of these effects indicates that large--scale magnetic features with length scales of order the pressure scale height will be produced by hydrodynamic instabilities close to the tachocline primarily at latitudes\footnote{In this section, we discuss locations in terms of the latitude, which is given by $90 \degrees-\theta$.} in the range $7 \degrees$--$30\degrees$. 

Active regions on the Sun primarily appear close to a central emergence latitude which varies from $\sim 30\degrees$ at the beginning of a solar cycle to $0 \degrees$ at its end. The observed spread in sunspot latitude is $\sim\pm 10 \degrees$ at any given time. If global dynamo effects prevent magnetic buoyancy from bringing coherent field structures to the surface far from the central emergence latitude, the number of active regions that can be produced by hydrodynamic modes in the tachocline will be determined by what portion of the latitude range $7 \degrees$--$30\degrees$ falls within $\sim\pm 10 \degrees$ of the central emergence latitude. Thus, we expect that there will be more active regions when the central emergence latitude is close to $15 \degrees$ than when it is near $0 \degrees$ or $30 \degrees$.  This is consistent with observations: solar maximum occurs when the central emergence latitude is $ \sim 15 \degrees$, while solar minimum occurs when the central emergence latitude is close to $0 \degrees$ or $30 \degrees$.

\subsection{MRI Saturation}\label{sec:saturation}

Even if the field is small as assumed in most of this work, $\omega_A \ll q\Omega$, when the MRI is initiated, that condition will not last long as the field grows exponentially. 
This growing field may be susceptible to tearing by associated thermal convection, 
but convection will be ineffective on large scales, in excess of the 
pressure scale height. The large--scale structure of the solar magnetic field will instead
depend on the interaction of local convective and MRI eddies with global dynamo effects, so the determination of the saturated field is nontrivial.  Nevertheless, it is useful to consider the saturated field produced by these individual effects separately.

In the absence of thermal convection, saturation of the MRI 
occurs for $v_{\rm A} \sim q \Omega r$ or $\omega_{\rm A} \sim q \Omega$
\citep{BH98,vishniac09}. 
That condition will be reached quickly, on the timescale $\Omega^{-1}$. While linear field winding that is responsible for the $\Omega$ effect in the dynamo context might be active in the tachocline, the MRI 
grows exponentially for $\theta\le 60 \degrees$ and may thus be responsible for the level of fields that are thought to 
then be driven to the solar surface by magnetic buoyancy. If so, the field winding would have no effect on the growth of field; it would only be responsible for creating large--scale toroidal fields from the strong small--scale poloidal fields produced by the MRI. The saturated field, $B_{\rm sat, MRI}$, resulting from the MRI is then
\begin{equation}
\label{eq:satmri}
 B_{\rm sat, MRI} \sim \sqrt{4 \pi \rho} |q| \Omega \varpi.
\end{equation} 

For a typical angular velocity 
of the tachocline, $\Omega_{\rm tach} = 2.7\times10^{-6}$ rad s$^{-1}$, 
and a density $\rho_{\rm tach} = 0.2$ \gcm at a spherical radius of 
$r_{\rm tach} \sim0.7~\rr$, and taking $\theta \sim 30 \degrees$ corresponding to a location of strong MRI growth, we find that the saturation value of the field produced by the MRI is $B_{\rm sat, MRI}\sim 1.2\times10^5 
|q|$ G. Since $q$ is of order unity in the tachocline, if the magnetic field near the solar surface originates in the tachocline, it could easily be produced by the MRI for $\theta\le 60 \degrees$.

In the absence of shear, the saturation field will be in equipartition with the turbulent pressure resulting from convection, so that $v_{\rm A} \sim {v_{\rm conv}}$, the convective velocity. Thus, the saturated magnetic field,  $B_{\rm sat, conv}$, resulting from convection will be 
\begin{equation}
\label{satconv}
 B_{\rm sat, conv} \sim \sqrt{4 \pi \rho}{v_{\rm conv}}.
\end{equation}
The convective velocities in the solar envelope were previously thought to range from about 0.05 \kms\ at the base of the convective zone to 
about 2.5 \kms\ at the solar surface \citep{CD96,Ho05}. Recent studies by \citet{hanasoge_seismic_2010} and \citet{hanasoge_anomalously_2012} have shown, however, that on large scales, the convective velocities are typically smaller than 0.01 \kms\ at large radii $r\sim 0.96 ~\rr$, with significantly larger velocities appearing only very close to the solar surface. For  $r= 0.96 ~\rr$, which corresponds to $\rho \sim 0.01$ \gcm, the saturated field resulting from convection is $ B_{\rm sat, conv} \sim 3.5\times 10^2$ G for $v_{\rm conv}\sim 0.01$ \kms\, and $ B_{\rm sat, conv} \sim 3.5\times 10^4$ G  for $v_{\rm conv}\sim 1$ \kms\. In either case, the saturation field, $B_{\rm sat, MRI}\sim 1.2\times10^5 $ G, that can be produced by the MRI in the tachocline at $\theta<60 \degrees$ and brought to the surface by magnetic buoyancy will dominate any field that 
is in equilibrium with the turbulent pressure resulting from convection near the surface.

Because convection is relatively weak except at $r>0.99 \rr$, MRI modes in the shear region close to the solar surface that was identified by \citet{M11} may be of greater importance than suggested by our linear analysis.  For $r=0.96 ~\rr$, $\theta=90 \degrees$,  $\Omega\sim2.5 \times 10^{-6}$ rad s$^{-1}$ and $\rho \sim 0.01$ \gcm, Equation (\ref{eq:satmri}) gives a magnetic field from the growth of MRI modes {\sl in situ} in the upper convection zone of $B_{\rm sat, MRI, is} \sim8\times10^4 |q| G$. Because $q\ltsim 1$ at this radius, the magnetic field produced {\sl in situ} by the MRI is comparable to the field that can be produced by convection except at $r>0.99 \rr$.  Thus, in addition to producing strong fields in the tachocline at $\theta\le 60 \degrees$, the MRI may play an important role in the origin of small-scale magnetic fields in the upper convection zone at all latitudes in the Sun.

\section{Conclusions} \label{sec:conclusions}
  \begin{deluxetable*}{llllll}
  \scriptsize
  \tablecolumns{6}
  \tablecaption{Table of Results}
\tablehead{\colhead{Region}& \colhead{Mode\,\tablenotemark{a}}     & \colhead{$\Gamma$\,\tablenotemark{b}}      &\colhead{Restrictions on $\mathbf{k}$\,\tablenotemark{c}}  &\colhead{Dependence on $\mathbf{B}$\,\tablenotemark{c} }&\colhead{Nonaxisymmetric?\,\tablenotemark{d}}}
\startdata 
TS & MRI (PSS)&$\sim \Omega$ & $|\Phi_{kB}|\approx90 \degrees$\ & $B_{\rm pol}>0.01$ G , $|\Phi_{Br}| \ltsim 45 \degrees$&Y        \\
   \nodata & MRI (CMRI) &$\sim 0.1 \Omega$&$|k_\varpi/k_Z|<1$ & $0.01$ G$<B_{\rm pol}<1 $ G& Y,  $B_{\rm pol}< 0.1$ G, $B_{\rm pol}>1$ G  \\
TU (in tachocline) &HS& $\sim \Omega$&$k_\varpi/k_Z\ll -1$&---&N\\
 \nodata                  &MRI& $\sim \Omega$&$k_\varpi/k_Z>-0.1$ &$B_{\rm pol}>0.01$ G &Y, $B_{\rm pol}< 0.1$ G     \\
TU (above tachocline)&C&$\sim \Omega$  &$|k_\varpi/k_Z|\gg1$  &---&N \\
        \nodata             &HS& $\sim\Omega$& $k_\varpi/k_Z\ll -1$ &---&N\\
               \nodata     &MRI& $\sim 0.1 \Omega$&$|k_\varpi/k_Z|<1$&$B_{\rm pol}>0.1$ G\tablenotemark{f}&Y, $B_{\rm pol}< 0.1$ G\tablenotemark{f}\\
                  \nodata  &C (overstable)\tablenotemark{e}&$\sim 0.1 \Omega$&$k_\varpi/k_Z<0$&---&N\\ 
TL & HS & $ \sim 0.1 \Omega$ & $k_\varpi/k_Z\gg1$&---&N\\
\nodata   & C  &  $\sim 0.1 \Omega$&$|k_\varpi/k_Z|\gg1$&---&N \\
C &C&$\sim |N|$&$|\Phi_{kr}|\neq 90 \degrees$&---&N
\enddata
\label{tab:conclusions}
\tablenotetext{a}{The three general types of mode are MRI modes with $\kdotva \sim \Omega$, hydrodynamic shear modes (HS) with $\kdotva \ll \Omega$, and convective modes (C) with $k^2 \xi \ll |N|$. When more than one mode of a general type exists, the specific mode designation is placed in parentheses. Modes are listed in order of descending growth rate.}
\tablenotetext{b}{These are approximate growth rates, rounded to the nearest order of magnitude.}
\tablenotetext{c}{These restrictions approximately define the regions of parameter space in which the axisymmetric mode can grow. A full specification  of the parameter space may be found by adding the constraints on $k$ and $\kdotva$ for the general mode type discussed in note (a).}
\tablenotetext{d}{This entry indicates with Y or N if the most rapidly growing mode of this type is nonaxisymmetric. If it is, approximate values of $B_{\rm pol}$ are given for which the nonaxisymmetric mode grows more quickly than the axisymmetric mode by a factor of 2 or greater.}
\tablenotetext{e}{The overstable convective mode has $k^2 \xi \sim |N|$ rather than following the condition $k^2 \xi < |N|$ typical for other convective modes.}
\tablenotetext{f}{These values are not discussed in the text; the differences between those given for the MRI mode in the tachocline part of Region TU are small and result from the differing strength of the shear at the two locations.}
\end{deluxetable*} 

In this paper, we have derived a dispersion relation for nonaxisymmetric instability including the MRI and used it to calculate the growth rate of modes throughout the Sun. We have explored the phase space defined by the magnitude and direction of the wavenumber ${\mathbf k}$, identifying the most rapidly growing modes at each location in the Sun. We have investigated the dependence of these growth rates on the initial magnitude and direction of the magnetic field ${\mathbf B}$. We have focused on the weak--field regime, corresponding to the initial formation of magnetic structures from a seed field.  We find that nonaxisymmetric effects typically represent a perturbation to the axisymmetric modes for toroidal--to--poloidal field ratios of $\sim 5$ that are typical in stellar field equilibria, so we first analyse the axisymmetric modes and then explore how the resulting conclusions are changed by nonaxisymmetric effects. Unless otherwise stated, specific numerical values in this section are based on an initial magnetic field with poloidal components oriented in the $r$ direction, toroidal--to--poloidal field ratio $R_{\rm TP}=5$, and poloidal field magnitude $B_{\rm pol}=0.2$ G.   Our conclusions, which are summarized in Table \ref{tab:conclusions}, are as follows:   
\begin{itemize}
\item The overall instability contains three types of submodes:  hydrodynamic convective modes, hydrodynamic shear modes, and small--scale MRI modes. The hydrodynamic modes are large--scale with wavelengths on the order of the pressure scale height, while the magnetohydrodynamic modes, which can be called MRI modes because they have $\kdotva \sim \Omega$ and are driven by shear, grow on much smaller length scales. The typical growth rates of the convective modes are on the order of the Brunt--V\"{a}is\"{a}l\"{a} frequency $N$, while the typical growth rates of the shear modes are typically $\ltsim \Omega$.

\item Those parts of the Sun in which significant growth of modes occurs may be divided into four regions in which the properties of these modes are significantly different: Region TS, which is located in the stably stratified tachocline, Region TU, which is located in the convectively unstable tachocline and lower convection zone at colatitudes $0\degrees < \theta < 60 \degrees$, Region TL, which is located in the convectively unstable tachocline and lower convection zone at colatitudes $\theta > 60 \degrees$, and Region C, which is located in the upper convection zone at all latitudes.  

\item The typical growth rates of the most rapidly growing modes are $\sim \Omega$ in Regions TS and at smaller radii in Region TU, $\sim 0.1 \Omega$ in Region TL, and $\sim N$ in Region C. These modes are driven by shear in Region TS and at the bottom of Region TU in the tachocline, by both shear and convection in the upper parts of Region TU and at the bottom of Region TL, and solely by convection in Region C and the upper parts of Region TL. In general, the importance of shear is greatest in the tachocline and typically decreases outward, while the strength of convection increases as one moves outward from the tachocline. 

\item In Region TS, all modes that grow significantly are small-scale MRI modes with $|\kdotva|\sim \Omega$ so that thermal diffusion can reduce the stabilization resulting from stable stratification, and all of these modes are driven by shear. Two types of MRI modes exist in this region: the CMRI mode with $|k_\varpi/k_Z|<1$ and typical growth rate $\sim 0.1 \Omega$ and the PSS mode with large $k$, 
$|\Phi_{kB}|\approx 90 \degrees$, and typical growth rate $\sim \Omega$.

\item In Region C, all modes are convective; their growth rate is $\sim |N|$ for all ${\mathbf k}$ such that $k^2 \xi <|N|$. The growth rate does not depend strongly on $k_\varpi/k_Z$ except at two locations: the growth of modes is entirely suppressed for $k_\varpi/k_Z\approx \tan \theta$, and is slightly enhanced for $k_\varpi/k_Z\approx -{\rm cot} \theta$. Convective flows are oriented in the $r$ direction; therefore, for $k_\varpi/k_Z\approx\tan \theta$, which corresponds to $\Phi_{kr}\approx 0 \degrees$, the perturbed velocity is perpendicular to $r$ and no growth occurs, while for $k_\varpi/k_Z\approx -{\rm cot} \theta$, which corresponds to $|\Phi_{kr}|\approx 90 \degrees$, the perturbed velocity is parallel to $r$, and the growth of instability is maximized. 

\item In Region TL, hydrodynamic shear and convective modes exist on large scales for $|k_\varpi/k_Z|\gg1$, so these modes drive flows in the $Z$ direction at nearly constant $\varpi$. Epicyclic stabilization, which is enhanced by the presence of positive radial shear in Region TL, prevents the growth of modes with $|k_\varpi/k_Z|<\tan \theta$ that cause flows to move in the $\varpi$ direction. This stabilization means that the growth rates of the most rapidly growing modes in Region TL are much smaller than those in any other region, typically smaller than $0.1 \Omega$.  Close to the equator at $\theta>83 \degrees$, this suppression is even more dramatic, with growth rates falling below $0.05 \Omega$ . The suppression of large--scale modes near the equator, combined with the suppression of large--scale modes by small--scale shear modes in Regions TS and TU suggested by \citet{PM07}, may explain why active regions tend to appear at latitudes of $\approx 20 \degrees$ in the Sun during solar maximum.  

\item In Region TU, we find that four modes are present, two of which are driven by convection and two of which are driven by shear. The two types of convective modes are a large--scale hydrodynamic mode of the type found in Region TL and a weak, highly overstable mode with $k^2 \xi \sim N$. The two types of shear mode are a large--scale hydrodynamic mode similar to that found in Region TL, and a small--scale shear mode with similar properties to that of the CMRI mode. In the upper tachocline, the shear modes grow much more quickly than the convective modes, and the most rapidly growing mode is the hydrodynamic shear mode. In contrast, in the lower convection zone, the most rapidly growing modes are large--scale hydrodynamic modes driven by both shear and convection, with convection being slightly more important.

\item The growth rate of the small--scale MRI modes has a significant dependence on the initial magnetic field strength and orientation. For the CMRI mode in Region TS, the magnetic field must be large enough that resistive and viscous dissipation are unimportant on scales for which $k^2 \xi >N$ but small enough that magnetic tension does not stabilize the growth of modes on these small scales. Therefore, at the bottom of Region TS, the growth of field occurs quickly only for a relatively small range of poloidal fields: $0.02$ G$<B_{\rm pol}<0.6$ G. For the PSS mode, the only restriction on the magnetic field is that $B>0.08$ G; for smaller fields, a hydrodynamic mode is present that does not depend on field orientation. Finally, for the small--scale mode in Region TU, the lack of stratification means that there is no upper limit to the magnetic field, as modes can grow on large scales; typically modes can grow for all $B_{\rm pol}>0.1$ G. Only the PSS modes in Region TS have a strong dependence on the field orientation. They only exist for fields that have $|\Phi_{Br}\ltsim 45 \degrees$, so that $|\Phi_{kB}|\approx90 \degrees$ corresponds to wavenumbers $\mathbf{k}$ that drive flows in the direction of the local shear, which is approximately the $r$ direction.  

\item Nonaxisymmetric small-scale MRI modes with $m\neq 0$ grow faster than their axisymmetric counterparts in Regions TS and TU. For the solar case, the explicitly nonaxisymmetric terms are typically unimportant, and the only important nonaxisymmetric effect is the influence of the normalized toroidal wavenumber $\mu$ on the value of $\kdotva$. The primary effect of nonaxisymmetry for CMRI modes and modes in Region TU is to break the degeneracy between $\kdotva$ and $k$, substantially increasing the growth rate for CMRI modes; nonaxisymmetric effects are significantly weaker for the PSS modes, because changes in $\mu$ can be balanced by adjustments in $k_\varpi/k_Z$.

\item The poloidal field magnitude $B_{\rm pol}$ has important effects on the relative growth rate of axisymmetric and nonaxisymmetric MRI modes. For the CMRI mode, nonaxisymmetric modes always have a growth rate at least twice that of the axisymmetric modes, but for very large and very small $B_{\rm pol}$, nonaxisymmetric modes are larger  by a factor of $>100$. For the PSS mode, nonaxisymmetric modes always have similar growth rates to axisymmetric growth modes. For the smal--scale shear mode in Region TU, nonaxisymmetric growth rates are similar to axisymmetric growth rates except for small $B_{\rm pol}$, where they can be larger by a factor of $>100$.

\end{itemize}

In this paper, we have applied a local WKB analysis to calculate the growth rates of modes. This technique has two important limitations that can be remedied in future work. The first limitation of this analysis is that the constraints on the value of the normalized toroidal wavenumber, $\mu$, implied by the local approximation have made it impossible to explore the full parameter space of nonaxisymmetric modes for small $R_{\rm TP}\sim 5$. An eigenvalue analysis of the full MRI would be an important next step that would make it possible to determine whether the dependence of growth rate on $\mu$ is the same for all toroidal fields.

The second and more serious shortcoming of the local WKB analysis is that it cannot explore the nonlinear effects of MRI modes.  The manner in which the physics of the MRI would affect the solar dynamo and observed field effects at the surface is a complex problem. 
From our analysis it seems that the physics of the MRI should 
be considered in this context in both the tachocline and in
the lower parts of the convective envelope itself. 

There is a general perception that the MRI is less important in 
stars than the Tayler--Spruit mechanism \citep{pt85,spruit99,spruit02} 
because the threshold of required shear for the latter in the radiative zone is lower than for the MRI.
While the MRI may be unimportant if shear grows very slowly, it is uncertain that this will be the case for all stars; if strong shear can be produced in a stellar model, the presence of diffusion may allow the MRI to grow quickly in radiative regions. Direct comparison between the two instabilities shows that the MRI grows 
exponentially rapidly on the timescale $\Omega^{-1}$ in any unstable
environment that has strong shear, while the typical timescale for the growth of the Tayler--Spruit instability is $\sim \Omega \waphi^{-2}$  for weak fields with $\waphi \ll \Omega$ \citep{pt85}.
Thus, if some environments in stars are unstable to the MRI, the growth rate for the MRI is likely to be much faster than that of
the Tayler--Spruit instability, by a factor of $(\Omega/ \waphi)^{2}$.  
\citet{PM07} have previously noted that the tachocline is probably also unstable to the
Tayler--Spruit mechanism, but argue that if both instabilities are 
present, the MRI is likely to dominate because of its much faster 
growth rate. It is also commonly assumed that magnetic nstabilities that depend on shear will not grow in convective zones \citep[e.g.,][]{heger_presupernova_2005}. MRI modes in convective zones are not hindered by stable stratification and can grow even when only moderate shear is present; because large--scale convective motions may be slow in convective regions, it is uncertain that the presence of convection will prevent MRI modes from growing, either. 

These considerations suggest that the role of the MRI in the evolution of stars in general deserves more attention. The saturation fields resulting from the MRI can be appreciable, 
and this may call for magnetic buoyancy effects to be re-evaluated. 
The MRI, as well as the Tayler--Spruit mechanism and other dynamo
processes, may also leave behind fossil fields in one stage of evolution 
that affect the physical conditions at later stages of evolution.
The role of magnetic fields in stellar evolution remains a major 
challenge requiring fully three--dimensional studies. This paper, investigating the combined effects of shear and convection
as well as nonaxisymmetric effects in the Sun, may constitute a guide in constructing these studies.

\acknowledgments
We thank the anonymous referee who greatly helped us to clarify our work in the context of that done in the solar community. We also thank Milos Milosavljevic, Pawan Kumar, Ethan Vishniac, Shizuka Akiyama, 
Swadesh Mahajan and Jesse Pino for helpful discussions and Rachel Howe
for providing the solar data. This work was begun at the Kavli Institute 
for Theoretical Physics; JCW is especially grateful for the supportive 
staff and conducive environment of KITP, which is supported by the National Science Foundation under Grant No. NSF PHY11-25915. Some work on this paper was 
also done in the hospitable environment of the Aspen Center for Physics,
which is supported by NSF Grant PHY-1066293. This work was supported in part 
by NSF Grants AST-0707769 and NSF AST-1109801. 
\\ \\ \\


\newpage
\begin{thebibliography}{}
\bibitem[Acheson(1978)]{acheson78} Acheson, D.~J.\ 1978, Royal 
Society of London Philosophical Transactions Series A, 289, 459 
\bibitem[Balbus(1995)]{bal95} Balbus, S.~A.\ 1995, \apj, 453, 380 
\bibitem[Balbus \& Hawley(1991)]{BH91} Balbus, S. A. \& Hawley, J. F.
 1991, \apj, 376, 214
\bibitem[Balbus \& Hawley(1994)]{BH94} Balbus, S.~A., \& Hawley, J.~F.\ 1994, \mnras, 266, 769 
\bibitem[Balbus \& Hawley(1998)]{BH98} Balbus, S. A. \& Hawley, J. F.  1998, Review of Modern Physics, 70, 1


\bibitem[Basu \& Antia(2001)]{BA01} Basu, S., \& Antia, H.~M.\ 2001, \mnras, 324, 498
\bibitem[Braithwaite(2009)]{Braith09} Braithwaite, J.\ 2009, \mnras, 397, 763 
\bibitem[Brown et al.(2011)]{Brown11} Brown, B.~P., Miesch, M.~S., 
Browning, M.~K., Brun, A.~S., \& Toomre, J.\ 2011, \apj, 731, 69
\bibitem[Buehler et 
al.(2013)]{buehler_quiet_2013} Buehler, D., Lagg, A., \& Solanki, S.~K.\ 2013, \aap, 555, A33 

\bibitem[Chanmugam(1979)]{Ch79} Chanmugam, G.\ 1979, \mnras, 187, 769 
\bibitem[Chandrasekhar(1960)]{chandra60} Chandrasekhar, S. 1960, Proc. Nat. Acad. Sci., 46, 253
\bibitem[Christensen-Dalsgaard et al.(1996)]{CD96} Christensen-Dalsgaard, J., 
    et al.\ 1996, Science, 272, 1286 


{bf \bibitem[Dikpati(2011)]{2011ApJ...733...90D} Dikpati, M.\ 2011, \apj, 733,  90 }

\bibitem[Dikpati \& Gilman(2001)]{dikpati_analysis_2001} Dikpati, M., \& Gilman, P.~A.\ 2001, \apj, 551, 536 

{bf \bibitem[Gilman  \& Fox(1997)]{1997ApJ...484..439G} Gilman, P.~A., \& Fox, P.~A.\ 1997, \apj, 484, 439 }

\bibitem[Hanasoge et al.(2010)]{hanasoge_seismic_2010} Hanasoge, S.~M., 
Duvall, T.~L., Jr., \& DeRosa, M.~L.\ 2010, \apjl, 712, L98 
\bibitem[Hanasoge et al.(2012)]{hanasoge_anomalously_2012} Hanasoge, S.~M., 
Duvall, T.~L., \& Sreenivasan, K.~R.\ 2012, Proc. Nat. Acad. Sci., 109, 11928 

\bibitem[Heger et al.(2005)]{heger_presupernova_2005} Heger, A., Woosley, 
S.~E., \& Spruit, H.~C.\ 2005, \apj, 626, 350 
\bibitem[Howe(2009)]{howe09} Howe, R.\ 2009, Liv. Rev. Sol. Phys., 6, 1 
\bibitem[Howe et al.(2005)]{Ho05} Howe, R., Christensen-Dalsgaard, J., 
Hill, F., Komm, R., Schou, J., \& Thompson, M.~J.\ 2005, \apj, 634, 1405 
\bibitem[Kim \& Ostriker(2000)]{KO00} Kim, W.-T., \& Ostriker, E.~C.\ 2000, \apj, 540, 372 
\bibitem[Maeder(2009)]{M09} Maeder, A.\ 2009, Physics, Formation and Evolution of Rotating Stars (Springer Berlin Heidelberg) 
\bibitem[Maeder \& Meynet(2005)]{MM05} Maeder, A., \& Meynet, G.\ 2005, \aap, 440, 1041 
\bibitem[Masada(2011)]{M11} Masada, Y.\ 2011, \mnras, 411, L26
\bibitem[Masada et al.(2007)]{MSS07} Masada, Y., Sano, T., Shibata, K.\ 2007, \apj, 655, 447
\bibitem[Masada et al.(2006)]{MST06} Masada, Y., Sano, T., Takabe, H.\ 2006, \apj, 641, 447 
\bibitem[Menou et al.(2004)]{MBS04} Menou, K., Balbus, S.~A., 
\& Spruit, H.~C.\ 2004, \apj, 607, 564

\bibitem[Ossendrijver(2003)]{OD03} Ossendrijver, M.\ 2003, \aapr, 11, 287
\bibitem[Parfrey \& Menou(2007)]{PM07} Parfrey, K.~P., \& Menou, K.\ 2007, \apjl, 667, L207 

\bibitem[Parker(1955)]{p55} Parker, E.~N.\ 1955, \apj, 122, 
293 

\bibitem[Paxton et al.(2011)]{paxton11} Paxton, B., Bildsten, 
L., Dotter, A., et al.\ 2011, \apjs, 192, 3 
\bibitem[Mart{\'{\i}}nez Pillet(2013)]{pillet_solar_2013} Mart{\'{\i}}nez 
Pillet, V.\ 2013, \ssr, 21 

\bibitem[Pitts \& Tayler(1985)]{pt85} Pitts, E., \& Tayler, R.~J.\ 1985, \mnras, 216, 139 
\bibitem[Rieutord et al.(2010)]{r10} Rieutord, M., Roudier, T., Rincon, F., et al.\ 2010, \aap, 512, A4 

{bf \bibitem[Spiegel  \& Zahn(1992)]{1992A&A...265..106S} Spiegel, E.~A., \& Zahn, J.-P.\ 1992, \aap, 265, 106 }

\bibitem[Spitzer(2006)]{spitzer06} Spitzer, L. Jr. 2006, The Physics of Fully Ionized Gases, New York: Dover 
\bibitem[Spruit(1999)]{spruit99} Spruit, H.~C.\ 1999, \aap, 349, 189 
\bibitem[Spruit(2002)]{spruit02} Spruit, H. C. 2002, \aap, 381, 923
\bibitem[{\v S}vanda(2013)]{sv13} {\v S}vanda, M.\ 2013, 
\apj, 775, 7 
\bibitem[Tassoul(1978)]{Tassoul78} Tassoul, J.-L.\ 1978, Princeton Series in 
physics, Princeton: University Press
\bibitem[Velikhov(1950)]{Vel59} Velikhov, E. P. 1959, J. Exp. Theoret. Phys. (USSR), 36, 1398
\bibitem[Vishniac(2009)]{vishniac09} Vishniac, E.~T.\ 2009, \apj, 696, 1021
\bibitem[Workman \& Armitage(2008)]{WA08} Workman, J.~C., \& Armitage, P.~J.\ 
   2008, \apj, 685, 406 

\end{thebibliography}
\end{document}